\documentclass[journal]{IEEEtran}
\usepackage{optidef}

\newcommand{\avec}{{\bf{a}}}

\newcommand{\evec}{{\bf{e}}}

\newcommand{\pvec}{{\bf{p}}}
\newcommand{\qvec}{{\bf{q}}}

\newcommand{\uvec}{{\bf{u}}}

\newcommand{\xvec}{{\bf{x}}}
\newcommand{\zvec}{{\bf{z}}}

\newcommand{\vvec}{{\bf{v}}}

\newcommand{\hvec}{{\bf{h}}}

\newcommand{\etavec}{{\bf{\eta}}}

\newcommand{\onevec}{{\bf{1}}}
\newcommand{\zerovec}{{\bf{0}}}

\newcommand{\muvec}{{\bf{\mu}}}
\newcommand{\alphavec}{{\bf{\alpha}}}
\newcommand{\Zeromat}{{\bf{0}}}

\newcommand{\Lambdamat}{{\bf{\Lambda}}}

\newcommand{\Amat}{{\bf{A}}}
\newcommand{\Bmat}{{\bf{B}}}
\newcommand{\Cmat}{{\bf{C}}}
\newcommand{\Dmat}{{\bf{D}}}
\newcommand{\Emat}{{\bf{E}}}

\newcommand{\Gmat}{{\bf{G}}}
\newcommand{\Hmat}{{\bf{H}}}

\newcommand{\Imat}{{\bf{I}}}
\newcommand{\Lmat}{{\bf{L}}}
\newcommand{\Mmat}{{\bf{M}}}
\newcommand{\Kmat}{{\bf{K}}}
\newcommand{\Umat}{{\bf{U}}}

\newcommand{\Smat}{{\bf{S}}}

\newcommand{\Rmat}{{\bf{R}}}
\newcommand{\Vmat}{{\bf{V}}}

\newcommand{\Xmat}{{\bf{X}}}
\newcommand{\Ymat}{{\bf{Y}}}

\newcommand{\define}{\stackrel{\triangle}{=}}

\newcommand{\Psimat}{\mbox{\boldmath $\Psi$}}

\def\btheta{{\mbox{\boldmath $\theta$}}}

\def\btheta{{\mbox{\boldmath $\theta$}}}

\def\betavec{{\mbox{\boldmath $\beta$}}}

\def\alphavec{{\mbox{\boldmath $\alpha$}}}

\def\gammavec{{\mbox{\boldmath $\gamma$}}}

\def\etavec{{\mbox{\boldmath $\eta$}}}

\def\thetavec{{\mbox{\boldmath $\theta$}}}

\def\muvec{{\mbox{\boldmath $\mu$}}}

\newcommand{\be}{\begin{equation}}
\newcommand{\ee}{\end{equation}}
\newcommand{\beqna}{\begin{eqnarray}}
\newcommand{\eeqna}{\end{eqnarray}}

\usepackage{amsmath,amssymb,amsthm,enumerate}
\usepackage{float}
\usepackage{mathtools}
\usepackage{multirow,subcaption,graphicx}
\usepackage{cite}
\usepackage{graphicx,nicefrac}
\usepackage{subcaption}
\graphicspath{ {./Figures/} }
\usepackage[ruled,vlined]{algorithm2e}
\usepackage{dsfont} 

\usepackage[printonlyused]{acronym}
\acrodef{mle}[MLE]{maximum likelihood estimator}
\acrodef{cmle}[CMLE]{constrained maximum likelihood estimator}
\acrodef{crb}[CRB]{Cram$\acute{\text{e}}$r-Rao Bound}
\acrodef{ccrb}[CCRB]{constrained Cram$\acute{\text{e}}$r-Rao Bound}
\acrodef{dlpf}[DLPF]{decoupled linear power flow}
\acrodef{ac}[AC]{alternating current}
\acrodef{dc}[DC]{direct current}
\acrodef{alm}[ALM]{augmented Lagrangian method}
\acrodef{admm}[ADMM]{alternating directional method of multipliers}
\acrodef{algoadmm}[ADMM]{Alternating Directional Method of Multipliers}
\acrodef{tls}[TLS]{total least squares}
\acrodef{pdf}[PDF]{probability density function}
\acrodef{fim}[FIM]{Fisher Information Matrix}
\acrodef{pmu}[PMUs]{phasor measurement units}
\acrodef{mse}[MSE]{mean-squared-error}
\acrodef{gsp}[GSP]{graph signal processing}
\acrodef{snr}[SNR]{signal-to-noise ratio}
\acrodef{gmrf}[GMRF]{Gaussian Markov random field}
\acrodef{glasso}[GLASSO]{graphical LASSO}
\acrodef{lgmrf}[LGMRF]{Laplacian-constrained GMRF}
\acrodef{i.i.d}[i.i.d.]{independent and identically distributed}
\acrodef{dn}[DN]{distribution network}
\acrodef{tn}[TN]{transmission network}
\DeclareMathAlphabet{\pazocal}{OMS}{zplm}{m}{n}

\SetKwInput{KwInput}{Input}
\SetKwInput{KwOutput}{Output}
\usepackage{bbm}
\usepackage{xcolor}

\ifCLASSINFOpdf

\else

\fi

\SetKwInput{KwInput}{Input}                
\SetKwInput{KwOutput}{Output}              
\SetKwInput{KwInit}{Initialization}        

\begin{document}

\title{Estimation of Complex-Valued Laplacian Matrices for Topology Identification in Power Systems
}
\author{\vspace{-0.15cm}Morad Halihal,
Tirza Routtenberg, \IEEEmembership{Senior Member, IEEE},
and  H. Vincent Poor \IEEEmembership{Fellow Member, IEEE}
\vspace{-0.25cm}

\thanks{
{\footnotesize{M. Halihal and T. Routtenberg are with the School of Electrical and Computer Engineering Ben-Gurion University of the Negev Beer-Sheva 84105, Israel, e-mail: moradha@post.bgu.ac.il, tirzar@bgu.ac.il.
T. Routtenberg and H. V. Poor
are with the Department of Electrical and Computer Engineering, Princeton University, Princeton, NJ, e-mail: \{tr4175,poor\}@princeton.edu.
}}}
\thanks{Part of this work, focusing on estimating admittance matrices under the AC model with separated sparsity constraints,  was presented at the IEEE International Conference on Acoustics, Speech, and Signal Processing (ICASSP) 2022 \cite{Morad_ICASSP2022}. This research was supported by the Israel Ministry of National
Infrastructure, Energy, and Water Resources and by an Electra scholarship.}
}
\maketitle
\begin{abstract}
In this paper, 
we investigate the problem of estimating a complex-valued Laplacian matrix with a focus on its application in the estimation of 
admittance matrices in power systems.
The proposed approach is based on a constrained maximum likelihood estimator (CMLE) of the complex-valued Laplacian, which is formulated as an optimization problem with  Laplacian and sparsity constraints. 
The complex-valued Laplacian is a symmetric, non-Hermitian matrix that exhibits a joint sparsity pattern between its real and imaginary parts.
{\textcolor{black}{Thus, we present a  group-sparse-based 
 penalized log-likelihood
approach for the Laplacian estimation.}}
Leveraging the {\textcolor{black}{mixed $\ell_{2,1}$ norm}} relaxation of the joint sparsity constraint,
we develop {\textcolor{black}{a new alternating direction method of multipliers (ADMM)}} estimation algorithm for the implementation of the CMLE of the Laplacian matrix {\textcolor{black}{under a linear Gaussian model}}.
Next, we apply  the proposed \textcolor{black}{ADMM} algorithms for the problem of estimating the admittance matrix under three commonly-used measurement models that stem from Kirchhoff’s and Ohm’s laws, each with different assumptions and simplifications: 
1)	the nonlinear alternating current (AC) model;
2)	the decoupled linear power flow (DLPF) model; and 
3)	the direct current (DC) model. 
The performance of the {\textcolor{black}{ADMM}}  algorithm is evaluated using data from  the IEEE $33$-bus power system data
 under different settings.
 The numerical experiments demonstrate that
the proposed algorithm outperforms existing methods in terms of mean-squared-error (MSE) and F-score, thus providing a more accurate recovery of the admittance matrix.

\end{abstract}
\section{Introduction}
The Laplacian matrix is a critical tool for studying the properties of graphs and is central to many key concepts in \ac{gsp}, machine learning, and networked data analysis.
Examples include semi-supervised learning, dimensionality reduction, community detection in complex networks, spectral clustering, and solving partial differential equations on graphs. It can also be used to define notions of smoothness and regularity on graphs, making it a powerful tool for analyzing and understanding the structure of complex data \cite{Lital2023Smooth, sandryhaila2014discrete, shuman2013emerging}. 
Thus, the estimation of the Laplacian matrix is a fundamental task in various fields.

Complex-valued graph signals arise in several real-world applications \cite{Amar_Routtenberg_2023}, such as multi-agent systems  \cite{han2015formation},
 wireless communication systems \cite{Tscherkaschin_2016},
 voltage and power phasors in electrical networks  \cite{wu2022complexvalue,drayer2020detection,dabush2021state,dinesh2023complex}, and  probabilistic graphical models with 
complex-valued multivariate Gaussian vectors \cite{Tugnait_2019}.
Despite the widespread use of complex-valued graph signals, 
 the recovery of complex-valued Laplacians is a crucial problem that has not been well explored.

Numerous approaches have been proposed in the literature for the estimation of the real-valued Laplacian matrix. 
Notably, the seminal work in \cite{10.1093/biostatistics/kxm045}  introduced a regularization framework for sparse precision (inverse covariance) estimation in graphical models,  giving rise to the \ac{glasso} algorithm.
Subsequently, researchers have explored various computationally efficient variations and extensions of \ac{glasso} in recent years (see, e.g.  \cite{10.5555, Mazumder2011TheGL}). 
In particular,  \ac{glasso} has recently been developed for proper and improper complex Gaussian graphical models in \cite{Tugnait_2019}.
Lately, there has been growing interest in developing estimation methods for cases where the precision matrix is a Laplacian matrix \cite{Egilmez_Pavez_Ortega_2017,ying2021minimax,Yakov2023Laplace}. 
The problem of learning the graph Laplacian from smooth graph signals has been considered in \cite{Xiaowen2015Laplace,Vassilis2016graph, Chepuri2016Laplace}.
However, these methods were developed for {\em{real-valued}} Laplacian matrices and for graphical models,  whereas our focus lies in the estimation of a {\em{complex-valued}}  Laplacian matrix in a physical linear Gaussian model, where the Laplacian matrix appears in the expectation.
This presents challenges, as the complex-valued Laplacian matrix is symmetric but not Hermitian, and exhibits a joint sparsity \cite{Golbabaee_Vandergheynst2012,NIPS2009_4ca82782} pattern between its real and imaginary parts.
Additionally, the solution of complex-valued optimization problems by using real-valued methods may be time-consuming and difficult, especially in the presence of complicated constraints \cite{hjørungnes2011complex}.

Admittance matrix estimation in power systems is a particular example of complex-valued estimation \textcolor{black}{that arises in both transmission and distribution networks \cite{Grotas_Routtenberg_2019, DekaDist2024, Park_Deka_Chertkov2018,CavaroInvProbing2019,Zhang2021, Swain2022}. This application 
is the main motivation of this work.} The modern electric grid is one of the largest and most complex cyber-physical systems today, where a power system can be represented as an undirected weighted graph with grid buses as nodes (vertices) and transmission lines as edges. Topology information plays a critical role in various aspects of power systems, including analysis, security, control, and stability assessment of power systems \cite{Abur_Gomez_book,Giannakis_Wollenberg_2013,7779155,8952865}.
Gaining knowledge of grid topology,  summarized in the admittance matrix, is essential for effective grid optimization and monitoring tasks. However, obtaining real-time information about topology and line parameters might be challenging, especially in distribution networks. Hence, accurate topology identification and precise line-parameter estimation become crucial for ensuring the efficiency and reliability of the network operations.
Additionally,  admittance matrix estimation can serve as a valuable tool for tasks such as identifying faults and line outages \cite{Poor_Tajer_2012}, and detection of potential cyberattacks \cite{drayer2020detection,Gal2023detection}. 
Furthermore, 
with the increasing integration of distributed renewable energy resources, the importance of admittance matrix estimation has risen significantly, making it an indispensable element in addressing the challenges of modern power systems.

Graph theory and graphical models have been utilized in power systems for various tasks (see, e.g,.   \cite{6374108,OPFZhu,A_Zamzam}). In particular, topology identification and the estimation of the topologies of distribution grids have been discussed in \cite{Soltan2017Zussman,7541005,Grotas_Routtenberg_2019, Deka_Chertkov_Backhaus_2017,Park_Deka_Chertkov2018}.
 Other works explored the detection of topological changes \cite{Emami_Abur_2010,Sharon_2012}  and the identification of edge 
disconnections \cite{shaked2021identification,Poor_Outage_identification} in electrical networks using hypothesis testing methods.
The methods in \cite{Li_Poor_Scaglione_2013,Anwar_Mahmood_Pickering_2016,Kekatos_Giannakis_Baldick2016,Xie_He2017} can reveal part of the grid information,  such as the grid connectivity and the eigenvectors of the topology matrix. Blind estimation of states and the real-valued admittance (susceptance) matrix has been studied in \cite{Grotas_Routtenberg_2019}. Additionally, 
in \cite{Park_Deka_Chertkov2018}, a novel algorithm based on the recursive grouping algorithm \cite{Choi2011M}, is suggested for learning the distribution grid topology and estimating transmission line impedances.
However, these methods estimate the {\em{real-valued}} admittance matrix and are often based on {\em{linearized}} power flow models (e.g. a \ac{dc} model), or on restrictive assumptions on the network structure (e.g. a tree graph).
Other methods rely on data-driven approaches using historical measurements. However, these approaches demand large amounts of data, which is often unavailable in current systems.
A nonlinear weighted least-square error algorithm is used to estimate line parameters in \cite{6966756}, while a total least-squares method
is applied in \cite{6672179}; 
however, the \ac{tls} method becomes unstable at low \ac{snr} values due to significant errors in power measurements.
To conclude, existing methods do not
estimate the complex-valued admittance matrix under Laplacian constraints using the full nonlinear \ac{ac} model.





 In this paper, we address the problem of estimating a complex-valued Laplacian matrix. The complex-valued Laplacian is a symmetric but non-Hermitian matrix, exhibiting almost joint sparsity patterns between its real and imaginary parts. This property necessitates the development of a unified algorithm capable of simultaneously estimating the two real-valued, sparse, Laplacian matrices under the joint-sparsity constraint.
 To recover the complex-valued Laplacian matrix, we propose a \ac{cmle} formulated as an optimization problem for a general objective function with graph Laplacian and joint-sparsity constraints. {\textcolor{black}{The proposed \ac{cmle} formulation is based on the $\ell_{2,0}$ mixed-norm, which captures the joint sparsity via the row sparsity \cite{steffens2016compact} of a matrix that consists of the off-diagonal elements of the real and imaginary parts of the Laplacian. }}
 To efficiently solve for the \ac{cmle}, we {\textcolor{black}{use a convex relaxation approach based on the
$\ell_{2,1}$ mixed-norm, which promotes group sparsity.
Then, we}} develop {\textcolor{black}{a new \ac{admm} \cite{Boyd-ADMM} approach for a linear quadratic model, stemming from a linear Gaussian observation model}}.
In the second part of this paper, we model the power network as a graph and describe the sparsity and Laplacian constraints of the admittance matrix. 
The joint sparsity property is experimentally validated for the admittance matrix in typical electrical networks.
We then apply the proposed estimation methods to three commonly-used power-flow measurement models: the nonlinear \ac{ac} model, the extended linear \ac{dlpf} model, and the linear \ac{dc} model, each with different assumptions and simplifications. 
We evaluate the performance of the proposed algorithm on IEEE 33-bus power system data under various settings and \ac{snr} values. Numerical experiments demonstrate that the proposed algorithm outperforms existing methods in terms of \ac{mse} and F-score, leading to a more accurate recovery of the admittance matrix.

{\em{Notation and Organization:}} 
In this paper, vectors are denoted by boldface lowercase letters and matrices by boldface uppercase letters.  The $K \times K$ identity matrix is denoted by $\Imat_{K}$, and $\onevec$ and $\zerovec$ denote the vectors of ones and zeros, respectively. 
The notations $|\cdot|$ and  $\otimes$ denote the determinant operator and the Kronecker product, respectively.   For any vector $\uvec$, $\|\uvec\|_{2}$ denotes the Euclidean norm.
For any matrix $\Amat$, 
$\Amat\succeq\Zeromat$ means that $\Amat$ is a positive semi-definite matrix,  and $\Amat^{-1}$ and $\|\Amat\|_{F}$ are its inverse
and Frobenius norm, respectively. The vector
${\text{Vec}}(\Amat)$ is obtained by stacking the columns of $\Amat$, \textcolor{black}{and similarly, $\text{Vec}_{\ell}(\Amat)$ consists of the elements of the lower triangular matrix  (without the diagonal) of a square matrix.}
 {\textcolor{black}{The $m$th element of the vector $\avec$ and
	the $(m,k)$th element of the matrix $\Amat$
		are denoted by $a_m$ and $[\Amat]_{m,k}$,  respectively. }}
 For a vector $\avec$, ${\text{diag}}(\avec)$ is a diagonal matrix whose $n$th diagonal entry is $a_n$; 
$\text{ddiag}(\avec)\define \text{diag}(\text{diag}(\avec))$,
 when applied to a matrix, and ${\text{diag}}(\Amat)$ is a vector with the diagonal elements of $\Amat$. In addition,
 ${\text{ddiag}}(\Amat)={\text{diag}}({\text{diag}}(\Amat))$.
 We also have
$x^{+}=\max\{x,0\}$.
 We denote by $\avec^i$ the $i$th row of the matrix $\Amat$. The
$(p, r)$ norm of a matrix $\Amat$ is defined as
\begin{equation}
    \label{definition of (r,p)-norm}
||\Amat||_{p,r}=\left(\sum\nolimits_{i=1}^M ||\avec^{(i)}||_p^r\right)^{\frac{1}{r}}.
\end{equation}
The gradient of a scalar function $f(\xvec)$ w.r.t. the vector $\xvec$ is a column vector with the dimensions of $\xvec$.

The rest of this paper is organized as follows. In Section \ref{section for task and problem formulation}, we describe the considered complex-valued Laplacian matrix estimation model, and discuss the properties of the resulting optimization problem. In Section \ref{Estimation Methods for General Objective}, we develop 
the \textcolor{black}{\ac{admm}} algorithm to solve this optimization. In Section \ref{models_section}, we implement the proposed algorithms for the three power system models.   Subsequently, simulations are shown in Section \ref{simulations}.
Finally, the paper is concluded in Section \ref{conclusions}.

\section{Complex-valued Laplacian estimation}
\label{section for task and problem formulation}
In this section, we formulate
the problem of estimating a complex-valued Laplacian matrix within general statistical models.
Since the primary focus of this paper lies in the application of admittance matrix identification, we derive the properties of the complex-valued Laplacian for linear quadratic models, as in electrical network theory. 
In particular, in  Subsection \ref{model_and_prob}, we represent the power network as a graph and describe the constraints imposed by the complex-valued Laplacian structure. In  Subsection \ref{CML_sub}, we formulate the constrained estimation problem of the complex-valued Laplacian for a general objective function. Finally, we discuss  the crucial constraint of joint sparsity between the real and imaginary parts of the Laplacian {\textcolor{black}{in Subsection \ref{joint_sparsity_subsec}.}}
\subsection{Model and problem formulation}
\label{model_and_prob}
A power system can be represented as an undirected connected weighted graph, ${\pazocal{G}}({\pazocal{V}},\xi)$, where
the set of vertices, ${\pazocal{V}}=\{1,\ldots,M\}$, comprises the buses (generators/loads) 
and the set of edges, $\xi$,  comprises the transmission lines.
The edge $(m,k)\in \xi$ corresponds to the transmission line between buses $m$ and $k$, and is characterized by the line admittance,
 $y_{m,k}\in {\mathbb{C}}$.
The system (nodal) admittance matrix  is a $M\times M$ complex
symmetric matrix, where its
 $(m,k)-th$ element 
  is given by
 (p. 97 in \cite{Monticelli1999})
\begin{equation}
\label{equ:YForm}
\displaystyle [\Ymat]_{m,k} = \begin{cases}
	\displaystyle \sum_{n\in{\mathcal{N}}_m }  y_{m,n},& m = k \\
  \displaystyle {\textcolor{black}{-}} y_{m,k},& m\neq k 
\end{cases},
\end{equation}
where $m,k=1,\ldots,M$.
{\textcolor{black}{It should be noted that the values of $y_{m,k}$ are assumed to be zero for nodes $m$ and $k$  that are not directly connected through a transmission line.}}
 In this graph representation, $\Ymat$ is a complex-valued Laplacian matrix.
 
The  admittance matrix from \eqref{equ:YForm}
can be decomposed as 
\be
\label{Y_def}
\Ymat=\Gmat + j\Bmat,
\ee
 where  the 
 conductance matrix, $\Gmat\in{\mathbb{R}}^{M\times M}$, and the  minus susceptance matrix, $\tilde{\Bmat}\define-\Bmat\in{\mathbb{R}}^{M\times M}$, are both {\textit{real-valued}} Laplacian matrices. That is,  $\tilde{\Bmat}$ and $\Gmat$ belong to
the set 
\beqna
\label{Lc}
    \mathcal{L}=\{\Lmat \in \mathcal{S}^M_+| [\Lmat]_{m,k}\leq 0, ~\forall m \neq k, \quad \Lmat\onevec=\zerovec\},
\eeqna
where $\mathcal{S}^M_+$ is the set of real, symmetric, positive semi-definite matrices.
As a result, it can be seen that $\Ymat$ is a symmetric but non-Hermitian matrix. Thus, its eigenvalues may not be real, and its eigenvectors may not be orthogonal \cite{Horn2012}.
For the sake of simplicity of computation, in this paper, we estimate the matrices $\Gmat$ and $\tilde{\Bmat}$, which are real-valued Laplacian matrices.

 Numerous Laplacian learning algorithms are based on the assumption that the Laplacian matrix is a sparse matrix. 
 In addition, for the considered application, electrical networks are known to be sparse, since the buses are usually only connected to 2-5 neighbors  \cite{Wang_Scaglione_Thomas_2010}. Therefore, we assume here that matrices $\Gmat$ and $\tilde{\Bmat}$ are sparse.

 To conclude, the real and imaginary parts of the complex-valued Laplacian matrix,  $\Gmat$ and $\tilde{\Bmat}$, belong to $ \mathcal{L}$ in \eqref{Lc} and are sparse. Thus, they  have the following properties:
	\renewcommand{\theenumi}{P.\arabic{enumi}}
	\begin{enumerate}
\item\label{P1}   Symmetry: $\Gmat=\Gmat^T$, $\tilde{\Bmat}=\tilde{\Bmat}^T$ 
\item\label{P2}     Null-space property: $\Gmat\onevec = \zerovec$, $  \tilde{\Bmat}\onevec = \zerovec$
\item\label{P4} Non-positive off-diagonal entries:  $[\Gmat]_{m,k}\leq 0$, \\ $[\tilde{\Bmat}]_{m,k}\leq 0$, $\forall   m,k=1,\dots,M$, $m\neq k$
\item\label{P3}    Positive semi-definiteness:  $\Gmat\succeq\zerovec$, $\tilde{\Bmat}\succeq \zerovec$ 
\item\label{P5} $\Gmat$ and $\tilde{\Bmat}$ are  sparse matrices.
\end{enumerate}

\subsection{\ac{cmle}}
\label{CML_sub}
The \ac{cmle} is a variant of the \ac{mle} that is used when there are parametric constraints on the unknown parameters to be estimated, and has appealing asymptotic performance \cite{Moore_scoring,Nitzan_constraints}.
In this work,
given a measurement model with the general {\textcolor{black}{negative}} log-likelihood $\psi(\Xmat;\Gmat,\tilde{\Bmat})$, our goal is to estimate $\Gmat$ and $\tilde{\Bmat}$ from the data, $\Xmat$.
The parametric constraints described by properties \ref{P1}-\ref{P5} of the Laplacian matrices $\Gmat$ and $\tilde{\Bmat}$ imply that the \ac{cmle} of $\Gmat$ and $\tilde{\Bmat}$ can be written as the following optimization problem:
\begin{equation}
\label{optimization_P1}
\begin{aligned}
&\min_{\Gmat,\tilde{\Bmat} \in \mathbb{R}^{M\times M}}\psi(\Xmat;\Gmat,\tilde{\Bmat}) \\
\textrm{s.t.}~& 
{\text{C.1}})~\Gmat=\Gmat^{T},~\tilde{\Bmat}=\tilde{\Bmat}^{T} \\ 
&{\text{C.2}})~\Gmat\onevec=\zerovec,~\tilde{\Bmat}\onevec=\zerovec\\
&{\text{C.3}})~[\Gmat]_{m,k}\leq 0,~ [\tilde{\Bmat}]_{m,k}\leq 0,~\forall m\neq k, k=1,\dots,M\\
&{\text{C.4}})~\Gmat\succeq\Zeromat,~\tilde{\Bmat}\succeq\Zeromat
\\
&{\text{C.5}})
~\Gmat,~\tilde{\Bmat} {\text{ are sparse matrices}}.
\end{aligned}
\end{equation}
It is known that symmetric, diagonally-dominant matrices with real nonnegative diagonal entries, such as  $\Gmat$ and $\tilde{\Bmat}$ under Constraints C.1-C.3, are necessarily positive semi-definite matrices (see, e.g. p. 392 in \cite{Horn2012}).  Consequently, Constraint C.4 is redundant and can be eliminated without affecting the solution of \eqref{optimization_P1}.

The estimation of the complex-valued Laplacian matrix in different applications involves the utilization of different measurement models. 
These models are functions of the network topology ($\Gmat$ and $\tilde{\Bmat}$) and the available measurements or given data ($\Xmat
$),  and may incorporate additional noise and errors. The specific {\textcolor{black}{negative}} log-likelihood of the measurement model,  $\psi(\Xmat,\Gmat,\Bmat)$, serves as the objective function in our optimization problem, as presented in \eqref{optimization_P1}. Notably, the constraints are identical across all models. 
This section provides a general overview of the problem, while a discussion of specific objective functions associated with different measurement models in power system applications is presented in Section \ref{models_section}.

\subsection{\textcolor{black}{Joint sparsity and Constraint C.5}}
\label{joint_sparsity_subsec}
Previous works have often assumed that the support sets of sparse matrices like $\Gmat$ and $\tilde{\Bmat}$ are either independent or identical \cite{Morad_ICASSP2022}. However, in real-world scenarios such as electrical networks, a more realistic assumption is that these matrices exhibit similar, but not identical, support sets. 
 In particular, {\textcolor{black}{in Subsection \ref{validation_subsec},}} we experimentally validate this property for the admittance matrix.
  This necessitates the development of a unified algorithm capable of simultaneously estimating the two real-valued, sparse, Laplacian matrices under the joint sparsity constraint \cite{Golbabaee_Vandergheynst2012,NIPS2009_4ca82782}.
{\textcolor{black}{Therefore,  Constraint C.5 in \eqref{optimization_P1} is replaced with the following row sparsity constraint}}:
\begin{equation}
\label{optimization_P1_alternative_form}
{\textcolor{black}{\tilde{\text{C}}}}.5)
~\left\|[\text{Vec}_{\ell}(\Gmat),\text{Vec}_{\ell}(\tilde{\Bmat})]\right\|_{p,0}\leq s, 
\end{equation}
where $s$ denotes the sparsity level. The $(p, 0)$ norm, as defined in \eqref{definition of (r,p)-norm} with $r=0$, counts the number of rows with nonzero entries, enforcing a row sparse structure. 
{\textcolor{black}{That is, in \eqref{optimization_P1_alternative_form} we define an $\frac{M(M-1)}{2}\times 2$ matrix, $[\text{Vec}_{\ell}(\Gmat),\text{Vec}_{\ell}(\tilde{\Bmat})]$, in which  each row contains the corresponding elements from the matrices $\Gmat$
 and $\tilde{\Bmat}$, e.g. the $(m, k)$th elements from both matrices. Then, 
 an inner $p$ norm is applied on the rows of this matrix, 
generating a vector of $p$ row-norms, i.e. a vector with the elements
\be
\label{vec_p}
\left( [\Gmat]_{m, k}^p + [{\tilde{\Bmat}}]_{m, k}^p \right)^{\frac{1}{p}}, ~~~m,k=1,\ldots,M,~~~ k< m.
\ee 
Finally, an outer zero semi-norm is applied to this resulting vector.
The inner $p$ norm provides a nonlinear coupling among the elements in a row (e.g. $[\Gmat]_{m,k}$
 and $[\tilde{\Bmat}]_{m,k}$), leading to the desired
row sparse structure of the matrix $[\text{Vec}_{\ell}(\Gmat),\text{Vec}_{\ell}(\tilde{\Bmat})]$.
}}
Consequently, applying this norm to the matrix $[{\text{Vec}}_{\ell}(\Gmat),{\text{Vec}}_{\ell}(\tilde{\Bmat})]$ {\textcolor{black}{and enforcing the constraint in \eqref{optimization_P1_alternative_form}},  implies that $\Gmat$ and $\tilde{\Bmat}$ share many joint zero elements, thereby adhering to the joint sparsity criterion. In addition, the operator $\text{Vec}_{\ell}$ returns the off-diagonal elements of its matrix argument that are below the diagonal.  As a result, the zero semi-norm constraint in \eqref{optimization_P1_alternative_form} is only applied to the below-diagonal elements of the matrices.

 {\textcolor{black}{The sparsity constraint,}} Constraint 
${\textcolor{black}{{\tilde{\text{C}}}}}.5$  {\textcolor{black}{on the $\ell_{p,0}$ norm}}, renders the optimization problem in \eqref{optimization_P1_alternative_form} non-convex {\textcolor{black}{and an NP-complete problem.
Thus,  convex relaxation in the form of an $\ell_{p,1}$ mixed-norm is considered in practice to obtain computationally tractable problems.}}
Common choices of the mixed-norms that promote group sparsity in statistical regression 
are the $\ell_{2,1}$ norm and the $\ell_{\infty,1}$ norm {\textcolor{black}{\cite{Yuan2006ModelSelection,Bach2008ConsistencyGroupLasso,Lv2011GroupLasso,Tugnait_2021}, commonly referred to as the group LASSO, that have been widely used for sparse learning of complex-valued signals.
Here we consider the $\ell_{2,1}$ norm, i.e. we use \eqref{definition of (r,p)-norm} with $p=2$ and $r=1$,  to promote the joint sparsity of the vectors $\text{Vec}_{\ell}(\Gmat)$ and $\text{Vec}_{\ell}(\tilde{\Bmat})$ and encourage simultaneous zero elements in $\Gmat$ and $\tilde{\Bmat}$.}}

Thereby, we transform the problem in \eqref{optimization_P1} into a convex optimization problem by removing Constraints C.4 {\textcolor{black}{(see the explanation after \eqref{optimization_P1})}} and ${\textcolor{black}{{\tilde{\text{C}}}}}.5$, and modifying the objective function in \eqref{optimization_P1} to obtain the following regularized convex optimization problem:
\color{black}\beqna
\label{optimization_P3}
\min_{\Gmat,\tilde{\Bmat} \in \mathbb{R}^{M\times M}} \psi(\Xmat,\Gmat,\tilde{\Bmat}) +\lambda \left\|[\text{Vec}_{\ell}(\Gmat),\text{Vec}_{\ell}(\tilde{\Bmat})]\right\|_{2,1}\nonumber \\
\textrm{s.t. C.1)-C.3) are satisfied},\hspace{2.25cm}
\eeqna
\color{black}
where $\lambda>0$ is a regularization parameter.
{\textcolor{black}{The last term in the objective function in \eqref{optimization_P3} can be interpreted as the $\ell_1$ norm applied on the vector defined in \eqref{vec_p}. Thus, this is a generalization of the classical $\ell_1$ norm regularization  \cite{tibshirani96regression,Kowalski2009SparseRegression}. }}
If  the unknown matrices are sufficiently sparse, and
under some conditions on  $N$ and on the model, the
minimization in \eqref{optimization_P3} is expected to approach the solution of \eqref{optimization_P1}.
{\textcolor{black}{It should be noted that even if $\psi(\Xmat,\Gmat,\tilde{\Bmat})$ is convex and differentiable, the objective function in \eqref{optimization_P3} is convex but non-smooth as the norm is non-differentiable. This is a consequence of the $\ell_{2,1}$ regularization term. This is similar to the classical Laplacian estimation in LASSO \cite{tibshirani96regression}, where the objective function is convex but not differentiable due to the presence of the $\ell_{1}$ regularization term.
In addition,}} it can be seen that the objective function in \eqref{optimization_P3} is not separable for the following reasons: 1) the first term of the objective function, $\psi(\Xmat,\Gmat,\tilde{\Bmat})$, may incorporate terms related to both $\Gmat$ and $\tilde{\Bmat}$; 2)  when choosing $p=2$ or $p=\infty$, the norm term is inseparable (as opposed to the $p=1$ case, {\textcolor{black}{which is inappropriate here since it does not enforce group sparsity)}}.
\vspace{-0.1cm}
\section{ADMM Method for General Objective}
\label{Estimation Methods for General Objective}
In this section, we discuss the joint estimation of the real-valued Laplacian matrices $\Gmat$ and $\tilde{\Bmat}$ by solving the regularized convex optimization problem presented in \eqref{optimization_P3}. 
\textcolor{black}{We focus on the \ac{admm}}, \textcolor{black}{presented in Subsection \ref{two_steps_ALM_subsection}},  which demonstrates promising efficiency and estimation performance. In Subsection \ref{remark_subsec}, \textcolor{black}{we discuss some issues regarding the proposed algorithm, including its computational complexity.}

\vspace{-0.12cm}
\subsection{\textcolor{black}{ADMM} algorithm}
\label{two_steps_ALM_subsection}
\textcolor{black}{In light of the above-mentioned considerations, we propose using the \ac{admm} \cite{Boyd-ADMM}, which is an efficient practical iterative approach for solving the regularized optimization problem in \eqref{optimization_P3}. The \ac{admm} is a powerful optimization technique that is well-suited for problems with non-differentiable convex objective functions and constraints. The \ac{admm} utilizes variable splitting; in our case, this implies introducing two additional auxiliary variables $\zvec_{G}$ and $\zvec_{B}$, in order to decouple the regularizers from the likelihood term $\psi(\Xmat,\Gmat,\tilde{\Bmat})$, and adds a consistency constraint. The resulting reformulation of \eqref{optimization_P3} is expressed as }
\beqna
\label{optimization_p4}
\min_{\Gmat,\tilde{\Bmat}\in\mathbb{R}^{M\times M},\textcolor{black}{\zvec_{G},\zvec_{B} \in \mathbb{R}^{\frac{M(M-1)}{2}}}} \hspace{-0.25cm}\psi(\Xmat,\Gmat,\tilde{\Bmat})  +\textcolor{black}{\lambda\|\bigr[\zvec_{G},\zvec_{B}\bigr]\|_{2,1}}\nonumber \\
\textrm{s.t.}~\left\{\begin{array}{l}
\textcolor{black}{{\text{C.0}})~\zvec_{G} =\text{Vec}_{\ell}(\Gmat),~\zvec_{B}=\text{Vec}_{\ell}(\tilde{\Bmat})}\\
\textrm{C.1)-C.3) are satisfied}.
\end{array}\right.
\eeqna
The scaled augmented Lagrangian 
that is used in the \textcolor{black}{\ac{admm}} approach
for the optimization problem in \eqref{optimization_p4}
is (p. 13-24, \cite{Boyd-ADMM})
\beqna
\label{Augmented_Lagrangian_function}
   &\psi(\Xmat,\Gmat,\tilde{\Bmat}) + \textcolor{black}{\lambda\|\bigr[\zvec_{G},\zvec_{B}\bigr]\|_{2,1}} \nonumber \\&\quad +  K_{\rho}(\Gmat,\textcolor{black}{\zvec_{G}},\muvec_{G},\Vmat_{G},\Lambdamat_{G},\textcolor{black}{\uvec_{G}})
   \nonumber
     \\&\quad +K_{\rho}(\tilde{\Bmat},\textcolor{black}{\zvec_{B}},\muvec_{B},\Vmat_{B},\Lambdamat_{B},\textcolor{black}{\uvec_{B}}) ,
\eeqna
where
 \begin{equation}
\label{K_p definition}
    \begin{aligned}
    &K_{\rho}(\Gmat,\zvec_{G},\muvec_{G},\Vmat_{G},\Lambdamat_{G},\uvec_{G}) \\&\quad
    \define \textcolor{black}{\frac{\rho}{2}(\|\zvec_{G}-\text{Vec}_{\ell}(\Gmat)+\rho^{-1}\uvec_{G}\|_{2}^2-\|\rho^{-1}\uvec_{G}\|_{2}^2)}
    \\&\quad +\frac{\rho}{2}(\|\Gmat\onevec+\rho^{-1}\muvec_{G}\|_{2}^2-\|\rho^{-1}\muvec_{G}\|_{2}^2)
    \\&\quad +\frac{\rho}{2}(\|\Gmat-\Gmat^{T}+\rho^{-1}\Vmat_{G}\|_{F}^2
    -\|\rho^{-1}\Vmat_{G}\|_{F}^2)
    \\&\quad+
    \frac{1}{2\rho}\sum_{m=1}^{M}\sum_{k=1,k\neq m}^{M}((\{[\Lambdamat_{G}]_{m,k}+\rho[\Gmat]_{m,k}\}^+)^2-[\Lambdamat_{G}]_{m,k}^{2}).
    \end{aligned}
\end{equation}
\textcolor{black}{The $K_{\rho}$ terms in \eqref{Augmented_Lagrangian_function} serve as penalty terms for the constraints associated with the two matrices. The first penalty term (first row of \eqref{K_p definition}) addresses Constraint C.0, and the subsequent terms (each one in a separate row of \eqref{K_p definition}) correspond to Constraints C.1, C.2, and C.3, respectively.
 In addition, 
 $\uvec_{G}$, $\uvec_{B}$, $\muvec_{G}$,
 $\muvec_{B}$, $\Vmat_{G}$, and $\Vmat_{B}$ are the scaled Lagrangian multipliers of the equality constraints (Constraints C.0-C.2), and $\Lambdamat_{G}$ and $\Lambdamat_{B}$ are the Lagrangian multiplier matrices of the inequality constraints (Constraint C.3). Finally, $\rho>0$ is called the penalty parameter. It should be noted that the term $K_{\rho}$ in \eqref{K_p definition} is obtained by substituting the constraints of the optimization problem from \eqref{optimization_p4} in the scaled form of the \ac{admm} (see, e.g. Subsection  3.1.1 on Page 15 of \cite{Boyd-ADMM}).}
 \textcolor{black}{
 Since the optimization problem in \eqref{optimization_p4} is convex,
the proposed \ac{admm} converges 
for all $\rho>0$ under mild conditions (Subsection 3.2 in \cite{Boyd-ADMM}).}

\textcolor{black}{The \ac{admm} algorithm has three main stages. In {\bf{Stage 1}},
the scaled augmented Lagrangian is minimized w.r.t. the 
decision variables, $\tilde{\Bmat}$ and $\Gmat$. Thus, based on \eqref{Augmented_Lagrangian_function}, in  this stage
we update the primal variables as follows:}
\beqna
    \label{Update G Augmented Lagrangian}
\Gmat^{(i+1)}=
\arg\min_{\Gmat \in \mathbb{R}^{M\times M}} 
\left\{\psi(\Xmat,\Gmat,\tilde{\Bmat}^{(i)})\right.\hspace{1.5cm}\nonumber
\\\left.+K_{\rho}(\Gmat,\textcolor{black}{\zvec_{G}^{(i)}},\muvec_{G}^{(i)},\Vmat_{G}^{(i)},\Lambdamat_{G}^{(i)},\textcolor{black}{\uvec_{G}^{(i)}})\right\},
\eeqna   
\beqna
    \label{Update B Augmented Lagrangian}
\tilde{\Bmat}^{(i+1)}=\quad
\arg\min_{\tilde{\Bmat} \in \mathbb{R}^{M\times M}} 
\left\{\psi(\Xmat,\Gmat^{(i+1)},\tilde{\Bmat})\right.\hspace{1.5cm}\nonumber
\\\left.+K_{\rho}(\tilde{\Bmat},\textcolor{black}{\zvec_{B}^{(i)}},\muvec_{B}^{(i)},\Vmat_{B}^{(i)},\Lambdamat_{B}^{(i)},\textcolor{black}{\uvec_{B}^{(i)}})\right\}.
\eeqna
In the following, we remove the iteration indices $(i)$ and $(i+1)$ from the different terms, for the sake of clarity.  
Since  the r.h.s. of 
 \eqref{Update G Augmented Lagrangian}-\eqref{Update B Augmented Lagrangian} is convex and differentiable (under the assumption that $\psi(\Xmat,\Gmat,\Bmat)$ is a convex and differentiable function)  w.r.t. $\text{Vec}(\Gmat)$ and $\text{Vec}(\Bmat)$, their minimum is attained by equating their derivatives w.r.t. 
 $\Gmat$ and $\Bmat$
 to zero, which results in 
\beqna
    \label{Update G equal zero}
\frac{\partial \psi(\Xmat,\Gmat,\tilde{\Bmat})}{\partial \text{Vec}(\Gmat)} +\frac{\partial K_{\rho}(\Gmat,\textcolor{black}{\zvec_{G}},\muvec_{G},\Vmat_{G},\Lambdamat_{G},\textcolor{black}{\uvec_{G}})}{\partial \text{Vec}(\Gmat)} = \zerovec,
\eeqna
and
\beqna
    \label{Update B equal zero}
\frac{\partial \psi(\Xmat,\Gmat,\tilde{\Bmat})}{\partial \text{Vec}(\tilde{\Bmat})}+\frac{\partial K_{\rho}(\tilde{\Bmat}, \textcolor{black}{\zvec_{B}},\muvec_{B},\Vmat_{B},\Lambdamat_{B},\textcolor{black}{\uvec_{B}})}{\partial \text{Vec}(\tilde{\Bmat})} = \zerovec.
\eeqna
It should be noted that in \eqref{Update G equal zero}-\eqref{Update B equal zero}, we applied a vec operator on both sides in order to have tractable terms. The resulting vectors of the derivatives are of size $M^2$.

\textcolor{black}{ 
In the models discussed in this paper, 
the function $\Psimat(\Xmat,\Gmat,\tilde{\Bmat})$ is a quadratic function w.r.t. $\Gmat$ and $\tilde{\Bmat}$  (see more details in Section \ref{models_section}), which enables us to simplify the expressions. Specifically, all the models discussed in Section \ref{models_section} can be expressed in the following general form:}
\begin{equation}
    \begin{aligned}
        \label{general_quadtratic_model}
        &\psi(\Xmat,\Gmat,\tilde{\Bmat})=\frac{1}{2}\sum_{n=0}^{N-1}\|
\avec_{1}+\Amat_{1}\Gmat\avec_{2}+\Amat_{2}\tilde{\Bmat}\avec_{3}\|_{2}^{2} \\&\quad \textcolor{black}{=  \frac{1}{2}\sum_{n=0}^{N-1}\|\avec_{1} + (\avec_{2}^T \otimes\Amat_{1}){\text{Vec}}(\Gmat)+(\avec_{3}^T\otimes \Amat_{2}){\text{Vec}}(\tilde{\Bmat})\|_{2}^{2}},
    \end{aligned}
\end{equation}
where $ \avec_{1}[n],\avec_{2}[n],\avec_{3}[n]\in\mathbb{C}^{M}$ and 
$\Amat_{1}[n],\Amat_{2}[n]\in\mathbb{C}^{M\times M}$ are all data vectors and matrices that are based on $\Xmat$. \textcolor{black}{(Note that due to space limitations and for the sake of clarity,  the dependency of these matrices and vectors on the time index $n$ is omitted)}.  That is, these matrices are determined by the specific observation model that is used, as discussed in Section \ref{models_section}. {\textcolor{black}{The last equality in \eqref{general_quadtratic_model} is obtained
by employing the vec operator properties and the Kronecker product.}}
\textcolor{black}{The general model in \eqref{general_quadtratic_model} is obtained in any case where the measurements can be represented as a linear function of  $\Gmat$ and $\tilde{\Bmat}$ with additive complex circularly symmetric Gaussian i.i.d  noise.
}
 
From \eqref{general_quadtratic_model}, we can obtain the partial derivatives of $\psi(\Xmat,\Gmat,\tilde{\Bmat})$ w.r.t. ${\text{Vec}}(\Gmat)$ and ${\text{Vec}}(\tilde{\Bmat})$ 
as follows:
\begin{equation}
    \label{derivative of Psi function w.r.t G}
    \begin{aligned}
        &\frac{\partial \psi(\Xmat,\Gmat,\tilde{\Bmat})}{\partial \text{Vec}(\Gmat)}= \Hmat_{1}\text{Vec}(\Gmat) + \Hmat_{2}\text{Vec}(\tilde{\Bmat}) + \hvec_{1},
    \end{aligned}
\end{equation}
\begin{equation}
    \label{derivative of Psi function w.r.t B}
    \begin{aligned}
        &\frac{\partial \psi(\Xmat,\Gmat,\tilde{\Bmat})}{\partial \text{Vec}(\tilde{\Bmat})} = \Hmat_{3}\text{Vec}(\Gmat)+\Hmat_{4}\text{Vec}(\tilde{\Bmat}) + \hvec_{2},
    \end{aligned}
\end{equation}
where 
\begin{subequations}
\be
\label{H1_H2_def}
\Hmat_i \define
\sum\nolimits_{n=0}^{N-1}\text{Re}\{(\avec_{2}^{*}\avec_{1+i}^{T}\otimes\Amat_{1}^{H}\Amat_{i})\},~~i=1,2,
\ee
\be
\label{H3_H4_def}\Hmat_i\define\sum\nolimits_{n=0}^{N-1}\text{Re}\{(\avec_{3}^{*}\avec_{i-1}^{T}\otimes\Amat_{2}^{H}\Amat_{i-2})\},
~~i=3,4,\ee
\be 
\label{h1_h2_def}
\hvec_i\define\sum\nolimits_{n=0}^{N-1}\text{Re}\{(\avec_{i+1}^{*}\otimes \Amat_{i}^{H})\avec_{1}\},~~i=1,2.
\ee
        \end{subequations}
        
The last two terms in \eqref{K_p definition}
are not a function of the diagonal elements of $\Gmat$. 
Thus,  the derivative of  \eqref{K_p definition} w.r.t. $[\Gmat]_{m,k}$ is given by
\begin{equation}
\label{draft 1}
    \begin{aligned}
 &\frac{\partial K_{\rho}(\Gmat,\textcolor{black}{\zvec_{G}},\muvec_{G},\Vmat_{G},\Lambdamat_{G},\textcolor{black}{\uvec_{G}})}{\partial[\Gmat]_{m,k}}
    \\ &=\bigg[(\rho\Gmat\onevec_{M}+\muvec_{G})\onevec_{M}^{T} + \rho(2\Gmat-2\Gmat^{T})\\&\quad+(\Vmat_{G}-(\Vmat_{G})^{T}))+
     {\mathds{1}}_{\{m\neq k\}}\bigg(\{\Lambda_{G}+\rho\Gmat\}^{+}\bigg) \\&\quad + \textcolor{black}{2(\text{Vec}^{T}(\Imat_{M})\otimes\Imat_{M})(\Imat_{M}\otimes(\Lmat_{\ell}(\Lmat_{\ell}^{T}\text{Vec}(\Gmat)-\zvec_{G}-\rho^{-1}\uvec_{G})}\bigg]_{m,k},
    \end{aligned}
\end{equation}
$\forall m,k=1,\dots,M$, \textcolor{black}{where we used the identity $\text{Vec}_{\ell}(\Gmat) = \Lmat_{\ell}^{T}\text{Vec}(\Gmat)$ (see Equation (2.140) in   \cite{hjørungnes2011complex}), and where $\Lmat_{\ell}\in \mathbb{R}^{M^2\times \frac{M(M-1)}{2}}$ is defined as follows:}
\textcolor{black}{\beqna
    \label{Permutation matrix Ll} 
     \Lmat_{\ell} \define \left[\text{Vec}(\evec_{2}\evec_{1}^{T}),\text{Vec}(\evec_{3}\evec_{1}^{T}),\dots,\right.\hspace{2.75cm}
     \nonumber\\\left. \text{Vec}(\evec_{M}\evec_{1}^{T}),\text{Vec}(\evec_{3}\evec_{2}^{T}),\dots,\text{Vec}(\evec_{M}\evec_{M-1}^{T})\right] .
\eeqna}

Now we can construct the update equation for $\Gmat^{(i+1)}$ by substituting both 
\eqref{derivative of Psi function w.r.t G} and  \eqref{draft 1}  into 
\eqref{Update G equal zero}. If we assume the case of $\{\Lambdamat_{G}+\rho\Gmat\}^{+}=\zerovec$, we obtain the following result:
\begin{equation}
    \begin{aligned}
    \label{G1 matrix}
            &\text{Vec}(\Gmat_{1}) = (\Hmat_{1}+\rho\Emat)^{-1} (-\Hmat_{2}\text{Vec}(\tilde{\Bmat})-\hvec_{1}-\gammavec_{G}),
    \end{aligned}
\end{equation}
where it is assumed that $\Hmat_{1}+\rho\Emat$ is a non-singular matrix,
\beqna
\label{Adef for G}
\textcolor{black}{\Emat}\define \onevec_{M}\onevec_{M}^{T}\otimes\Imat_{M} + \textcolor{black}{2}\Imat_{M^2}
\hspace{2.5cm}\nonumber\\
-2\sum_{i=1}^{M}\sum_{j=1}^{M}(\evec_{i}\evec_{j}^{T})\otimes(\evec_{j}\evec_{i}^{T}) + \textcolor{black}{\Lmat_{\ell}\Lmat_{\ell}^{T}},
\eeqna
and
\begin{equation}
\label{gamma_def}
\gamma_{G}\define \text{Vec}(\muvec_{G}\onevec_{M}^{T}+(\Vmat_{G}-(\Vmat_{G})^{T})  \textcolor{black}{-\Lmat_{\ell}(\rho\zvec_{G}+\uvec_{G})}).
\end{equation}
Conversely, under the assumption $\{\Lambdamat_{G}+\rho\Gmat\}^{+}=\Lambdamat_{G}+\rho\Gmat$, we obtain
    \begin{eqnarray}
    \label{G2 matrix}
            \text{Vec}(\Gmat_{2}) = (\Hmat_{1}+\rho(\Emat+\Imat_{M^{2}}))^{-1}\hspace{2.5cm}\nonumber\\ \times (-\Hmat_{2}\text{Vec}(\tilde{\Bmat})-\hvec_{1}-\gammavec_{G}-\text{Vec}(\Lambdamat_{G})).
\end{eqnarray}
Combining these two scenarios per element, i.e. using \eqref{G1 matrix} for element $(m,k)$ such that 
$[\Lambdamat_{G}+\rho\Gmat]_{m,k}^{+}=0$ and \eqref{G2 matrix} for element $(m,k)$ such that 
$[\Lambdamat_{G}+\rho\Gmat]_{m,k}>0$, 
we can formulate the complete update equation for $\Gmat^{(i+1)}$ as 
\begin{equation}
    \begin{aligned}
    \label{Update equation for G final}
        &\Gmat = \Mmat_{G} \odot \Gmat_{1} + (\onevec_{M}\onevec_{M}^{T} - \Mmat_{G}
        ) \odot \Gmat_{2} 
        ,
    \end{aligned}
\end{equation}
where
\begin{equation}
\begin{aligned}
\label{MG}
[\Mmat_{G}]_{m,k} &= \begin{cases} 
1 & \text{if } [\Lambdamat_{G} + \rho \Gmat_{1}]_{m,k} \leq 0 
\\
1 & \text{if } m=k\\
0 & \text{otherwise} 
\end{cases},
\end{aligned}
\end{equation}
and $\Gmat_{1}$ is defined in \eqref{G1 matrix}.
 The indicator matrix $\Mmat_{G}$ in \eqref{MG} distinguishes between the elements of the estimation of the matrix $\Gmat$ that have positive and negative values. In addition, it should be noted that in each iteration, the terms on the r.h.s. of \eqref{Update equation for G final}
are the current values of the parameters, while on the l.h.s. we obtain the updated estimation $\Gmat^{(i+1)}$.

Due to the symmetry of the derivative of the scaled augmented Lagrangian function in \eqref{Augmented_Lagrangian_function} w.r.t. $\Gmat$ and $\Bmat$, we can derive the update equation for $\tilde{\Bmat}^{(i+1)}$ using a similar procedure as for
$\Gmat$ in \eqref{Update equation for G final}.
 Therefore, under the assumption that $\Hmat_{4}+\rho\Emat$ is non-singular, the update equation for $\tilde{\Bmat}$ is
\begin{equation}
\label{update equation for B step - general}
\begin{aligned}
&\tilde{\Bmat} = \Mmat_{B}\odot\tilde{\Bmat}_{1} + (\onevec_{M}\onevec_{M}^{T}-\Mmat_{B}
)\odot\tilde{\Bmat}_{2}^{(i+1)} 
,
\end{aligned}
\end{equation}
where  
\begin{equation}
\label{update equation for B_1 step - general}
\text{Vec}(\tilde{\Bmat}_{1}) = (\Hmat_{4}+\rho\Emat)^{-1}(-\Hmat_{3}\text{Vec}(\Gmat)-\hvec_{2} -\gammavec_{B}),
\end{equation}
\begin{eqnarray}
\label{update equation for B_2 step - general}
    \text{Vec}(\tilde{\Bmat}_{2}) = (\Hmat_{4}+\rho(\Emat+\Imat_{M^2}))^{-1}\hspace{2.5cm}\nonumber\\ \times(-\Hmat_{3}\text{Vec}(\Gmat)-\hvec_{2}-\gammavec_{B}-\text{Vec}(\Lambdamat_{B})),
\end{eqnarray}
 $\Emat$ is defined as in \eqref{Adef for G}, and $\gammavec_{B}$ is defined as $\gamma_{G}$ in \eqref{gamma_def}, where we replace $\muvec_{G},\Vmat_{G}$,  \textcolor{black}{$\uvec_{G}$, and $\zvec_{G}$} with $\muvec_{B},\Vmat_{B}$, \textcolor{black}{$\uvec_{B}$, and $\zvec_{B}$}, respectively.
Finally, the indicator matrix in this case is 
\begin{equation}
\begin{aligned}
[\Mmat_{B}]_{m,k} &= \begin{cases}
1 & \text{if } [\Lambdamat_{B} + \rho \tilde{\Bmat}_{1}]_{m,k} \leq 0 \\
1 &m = k\\
0 & \text{otherwise}
\end{cases},
\end{aligned}
\end{equation}
which is similar to $\Mmat_{G}$, and distinguishes between the elements of the estimation of the matrix $\Bmat$ that have positive and negative values, i.e.
provides the condition for selecting the appropriate elements from $\tilde{\Bmat}_{1}$ and $\tilde{\Bmat}_{2}$ during the matrix update in \eqref{update equation for B step - general}.
Again, in each iteration, the terms on the r.h.s. of \eqref{update equation for B step - general}
are the current values of the parameters, while on the l.h.s. we obtain the updated estimation $\Bmat^{(i+1)}$.

\textcolor{black}{In \textbf{Stage 2}, the z-minimization step, the updates of  $\zvec_{G}$ and $\zvec_{B}$ are obtained by minimizing the scaled augmented Lagrangian from \eqref{Augmented_Lagrangian_function} 
w.r.t. $\zvec_{G}$ and $\zvec_{B}$, after updating the matrices $\Gmat$ and $\tilde{\Bmat}$ to 
$\Gmat^{(i+1)}$ and $\tilde{\Bmat}^{(i+1)}$, which results in
\begin{equation}
\label{Updating Z_G joint sparse}
\displaystyle [\zvec_{G}]_{m} = \begin{cases}
	\displaystyle \frac{\|\avec^{(m)}\|_{2}-\frac{\lambda}{\rho}}{\|\avec^{(m)}\|_{2}}\avec_1^{(m)}  ,& \|\avec^{(m)}\|_{2}>\frac{\lambda}{\rho} \\
  \displaystyle  0,& \text{otherwise} \\
\end{cases},
\end{equation}
\begin{equation}
\label{Updating Z_B joint sparse} 
\displaystyle [\zvec_{B}]_{m} = \begin{cases}
	\displaystyle \frac{\|\avec^{(m)}\|_{2}-\frac{\lambda}{\rho}}{\|\avec^{(m)}\|_{2}}[\avec^{(m)}]_{2}  ,& \|\avec^{(m)}\|_{2}>\frac{\lambda}{\rho} \\
  \displaystyle  0,& \text{otherwise} \\
\end{cases},
\end{equation}
 where  $\avec^{(m)}$ is the m-th row in the matrix $\Amat\define\left[\text{Vec}_{\ell}(\Gmat)-\rho^{-1}\uvec_{G},\text{Vec}_{\ell}(\tilde{\Bmat})-\rho^{-1}\uvec_{B}\right]$}. 
 
In \textbf{Stage 3} of the \textcolor{black}{\ac{admm}, which consists of the dual variables update,} we update the Lagrangian multipliers by employing gradient ascent. This involves computing the gradient of the scaled augmented Lagrangian function in \eqref{Augmented_Lagrangian_function} w.r.t. each Lagrange multiplier individually, which results in
\begin{align}
\label{multipliers augmented lagrangian}
\uvec_{G}^{(i+1)} 
&= \uvec_{G}^{(i)} + \rho\left(\zvec_{G}^{(i+1)}-\text{Vec}_{\ell}(\Gmat^{(i+1)})\right),\nonumber\\
\uvec_{B}^{(i+1)}
&= \uvec_{G}^{(i)} + \rho\left(\zvec_{B}^{(i+1)}-\text{Vec}_{\ell}(\tilde{\Bmat}^{(i+1)})\right),\nonumber\\
\muvec_{G}^{(i+1)}
&= \muvec_{G}^{(i)}+\rho\Gmat^{(i+1)}\onevec,\nonumber\\
\muvec_{B}^{(i+1)}
&= \muvec_{B}^{(i)}+\rho\tilde{\Bmat}^{(i+1)}\onevec,\nonumber\\
\Vmat_{G}^{(i+1)}
&= \Vmat_{G}^{(i)}+\rho\left(\Gmat^{(i+1)}-(\Gmat^{(i+1)})^{T}\right),\nonumber\\
\Vmat_{B}^{(i+1)}
&= \Vmat_{B}^{(i)}+\rho(\tilde{\Bmat}^{(i+1)}-(\tilde{\Bmat}^{(i+1)})^{T}),\nonumber\\
\Lambdamat_{G}^{(i+1)}
&= \{\Lambdamat_{G}^{(i)}+\rho\Gmat^{(i+1)}\}^{+},\nonumber
\\
\Lambdamat_{B}^{(i+1)}
&= \{\Lambdamat_{B}^{(i)}+\rho\tilde{\Bmat}^{(i+1)}\}^{+}.
\end{align}

The \textbf{stopping criterion} for the algorithm is based on the maximal number of iterations or on the convergence of the estimators, when $\|\Gmat^{(i+1)}-\Gmat^{(i)}\|$ and $\|\Bmat^{(i+1)}-\Bmat^{(i)}\|$ are small enough.
The algorithm is summarized in Algorithm  \ref{ALGO2}.

\begin{algorithm}[hbt]
\SetAlgoLined
\KwInput{\begin{itemize} 
   \item $M$ - Number of nodes
   \item  $\Xmat$ - data 
   \item $N_{max}$ - maximal number of iterations
   \item $\epsilon>0$ - convergence parameter
   \item $\rho>0$ - penalty parameter
   \end{itemize}}
   \KwOutput{$\hat{\Gmat}=\Gmat^{(i)}$, $\hat{\Bmat}=\Bmat^{(i)}$}
  \KwInit{
\begin{itemize}
\item Set $i=0$.
\item Initialize $\Gmat^{(0)}$ and $\Bmat^{(0)}$ using \textbf{Algorithm \ref{ALGO3}}.
\item Initialize $\Umat_{G}^{(0)}$, $\Umat_{B}^{(0)}$, $\Vmat_{G}^{(0)}$, $\Vmat_{B}^{(0)}$, $\Lambdamat_{G}^{(0)}$, and $\Lambdamat_{B}^{(0)}$ to the zero matrices of size $M \times M$.
\item Initialize $\muvec_{G}^{(0)}=\zerovec_M$ and $\muvec_{B}^{(0)}=\zerovec_M$.
\item \textcolor{black}{Calculate $(\Hmat_{1} + \rho \Emat)^{-1}$ and $(\Hmat_{1} + \rho (\Imat_{M^{2}}+\Emat))^{-1}$ using \eqref{H1_H2_def} and \eqref{Adef for G}.}
\item \textcolor{black}{Calculate $(\Hmat_{4} + \rho \Emat)^{-1}$ and $(\Hmat_{4} + \rho (\Imat_{M^{2}}+\Emat))^{-1}$ using \eqref{H3_H4_def} and \eqref{Adef for G}.}
\end{itemize}
}
\For{$i=1,\ldots,N_{max}$
}{
\begin{itemize}
\item  Update $\Gmat^{(i+1)}$ using \eqref{Update equation for G final}.
\item Update  
$\tilde{\Bmat}^{(i+1)}$ using \eqref{update equation for B step - general}.
\item \textcolor{black}{Update $\zvec_{G}^{(i+1)}$ and $\zvec_{B}^{(i+1)}$ using \eqref{Updating Z_G joint sparse} and \eqref{Updating Z_B joint sparse}}.
\item Update the Lagrange multipliers using \eqref{multipliers augmented lagrangian}.
\item Update $i\leftarrow i+1$
\item \If{ $\|\Gmat^{(i+1)}-\Gmat^{(i)}\|_{\textbf{F}}^{2}<\epsilon$ and $\|\tilde{\Bmat}^{(i+1)}-\tilde{\Bmat}^{(i)}\|_{\textbf{F}}^{2}<\epsilon$}{break}
\end{itemize}
}
\caption{\textcolor{black}{\ac{admm}} algorithm for joint sparse topology estimation}\label{ALGO2}
\end{algorithm}

The following are the initialization and final steps of the algorithm.
\subsubsection{Initialization}
The \textcolor{black}{\ac{admm}} in Algorithm \ref{ALGO2} requires a feasible starting point for the Laplacian matrices. We propose an initialization algorithm, outlined in Algorithm \ref{ALGO3}, which is based on the real and imaginary parts of the sample covariance matrices. These covariance matrices are symmetric (satisfying Constraint C.1) and positive semi-definite (satisfying Constraint C.4). We adjust their off-diagonal elements to be non-positive (to satisfy Constraint C.3), and modify their diagonal elements such that the sum of each row/column is 0 (to satisfy Constraint C.2). This initialization step ensures that the output, $\Gmat^{(0)}$ and $\tilde{\Bmat}^{(0)}$, are real-valued Laplacian matrices.
\begin{algorithm}[hbt]
\label{alg:the_alg}
\KwInput{\begin{itemize} 
   \item $M$ - Number of nodes
   \item  $\Xmat=[\xvec[0],\ldots,\xvec[N-1]]$ - complex-valued data
   \end{itemize}}
   \KwOutput{$\Gmat^{(0)}$ and $\tilde{\Bmat}^{(0)}$}
 \begin{itemize}
\item Calculate the real-valued sample covariance matrices
\end{itemize}
\beqna
\Smat_G = \frac{1}{N}\sum\nolimits_{n=0}^{N-1}\left(\text{Re}\{\xvec[n]\}-\frac{1}{N}\sum\nolimits_{m=0}^{N-1}\text{Re}\{\xvec[m]\}\right)
\nonumber\\
\times \left(\text{Re}(\xvec[n])-\frac{1}{N}\sum\nolimits_{m=0}^{N-1}\text{Re}\{\xvec[m]\}\right)^{T} \nonumber
\eeqna
\beqna
\Smat_B = \frac{1}{N}\sum\nolimits_{n=0}^{N-1}\left(\text{Im}\{\xvec[n]\}-\frac{1}{N}\sum\nolimits_{m=0}^{N-1}\text{Im}\{\xvec[m]\}\right)
\nonumber\\
\times \left(\text{Im}\{\xvec[n]\}-\frac{1}{N}\sum\nolimits_{m=0}^{N-1}\text{Im}\{\xvec[m]\}\right)^{T} \nonumber
\eeqna
\begin{itemize}
\item Remove the positive off-diagonal elements of $\Smat_G$, $\Smat_B$:
    \\
    $\Gmat^{(0)} = \text{ddiag}(\Smat_G)-\left\{-\Smat_G+\text{ddiag}(\Smat_G)\right\}^+$
    \\
        $\tilde{\Bmat}^{(0)} = \text{ddiag}(\Smat_B)-\left\{-\Smat_B+\text{ddiag}(\Smat_B)\right\}^+$
    \item Generate matrices that satisfy the null-space property:
    $\Gmat^{(0)} = \Gmat^{(0)}-\text{diag}(\Gmat^{(0)}\onevec_{M})$
    \\
    $\tilde{\Bmat}^{(0)} =\tilde{\Bmat}^{(0)}-\text{diag}(\tilde{\Bmat}^{(0)}\onevec_{M})$
    \end{itemize}
        \caption{Initialization of the matrices $\Gmat$ and $\tilde{\Bmat}$}\label{ALGO3}
\end{algorithm}


\subsubsection{Post-optimization adjustments}
\label{post_subsec}
After the termination of Algorithm \ref{ALGO2}, we 
apply a transformation approach
to ensure that the output of the algorithm satisfies the imposed constraints.
 First, we map $\hat{\Gmat}$ and $\hat{\Bmat}$ onto the subspace of symmetric matrices (Constraint C.1) using the  transformation
\begin{equation}
\label{projecting G to symmetric subspace}
    \begin{aligned}\hat{\Gmat}=\frac{\hat{\Gmat}+\hat{\Gmat}^{T}}{2}
~~{\text{and }}~
    \hat{\Bmat}=\frac{\hat{\Bmat}+\hat{\Bmat}^{T}}{2}.\nonumber
    \end{aligned}
\end{equation}
Second, to satisfy Constraint C.2, we replace the diagonal of $\hat{\Gmat}$  by ${\text{diag}}(\hat{\Gmat})-\hat{\Gmat}\onevec_{M}$, and similarly, the diagonal elements of $\hat{\Bmat}$ are replaced by ${\text{diag}}(\hat{\Bmat})=\text{diag}(\hat{\Bmat})-\hat{\Bmat}\onevec_{M}$.
Third, for Constraint C.3, we perform the mapping as follows: 
\begin{equation}
\label{projecting G to C.3}
    [\hat{\Gmat}]_{m,k}=-\{-[\hat{\Gmat}]_{m,k}\}^+,
{\text{ and }}
    [\hat{\Bmat}]_{m,k}=-\{-[\hat{\Bmat}]_{m,k}\}^+,\nonumber
\end{equation}
$\forall m\neq k$,  $m,k=1,\dots,M$.
Finally, we threshold the off-diagonal elements in the result. We set the threshold $\tau_{G}$ and $\tau_{B}$ to be smaller than the magnitude of the smallest estimated element of the diagonal, i.e. to
\begin{equation}
\label{Definition of tau_G}
    \begin{aligned}
    \tau_G\define\frac{1}{M}\cdot\min_{m=1,\dots,M} \hat{\Gmat}_{m,m},
~
    \tau_B\define\frac{1}{M}\cdot\min_{m=1,\dots,M} \hat{\Bmat}_{m,m}.\nonumber
    \end{aligned}
\end{equation}
This thresholding step promotes the sparsity of the solution.

\vspace{-0.25cm}
\subsection{\textcolor{black}{Remarks}}
\label{remark_subsec}
\subsubsection{Computational complexity}
\label{comp_subsec}
The analysis of the computational complexity of the proposed \textcolor{black}{\ac{admm}} is based on the number of matrix multiplications required per iteration.
The inversion of an $M^{2} \times M^{2}$ matrices in \eqref{G1 matrix}, \eqref{G2 matrix}, \eqref{update equation for B_1 step - general}, and \eqref{update equation for B_2 step - general} has a computational cost of $\mathcal{O}(M^{6})$ operations.
\textcolor{black}{ However, it does not directly impact the per-iteration computational cost since it can be performed offline, and thus can be precomputed or considered as input to the algorithm.}
\textcolor{black}{These matrices are only used for matrix-vector multiplications and are not to be calculated at each iteration.}
Within each iteration, the algorithm involves a matrix multiplication of precomputed matrices with a vector of size $M^2$, which has a computational cost of $\mathcal{O}(M^4)$ operations. Additionally, the algorithm utilizes a Hadamard product and summation operation (in \eqref{Update equation for G final}, \eqref{update equation for B step - general}). The Hadamard product between matrices of size $M \times M$ requires $\mathcal{O}(M^{2})$ operations, and the subsequent summation operation also has a computational cost of $\mathcal{O}(M^{2})$ operations. \textcolor{black}{The z-minimization step requires $\mathcal{O}(M^{2})$ operations}, and the update of the Lagrangian multipliers \eqref{multipliers augmented lagrangian} has a computational cost of $\mathcal{O}(M^3)$.
Given the precomputation of matrix inversions, the dominant computational cost for each iteration of the \textcolor{black}{\ac{admm}} algorithm is of the order of $\mathcal{O}(M^4)$. 

\color{black}
\subsubsection{Relation with Laplacian constrained GLASSO}
\label{glasso_subsec}

The \ac{glasso} \cite{10.1093/biostatistics/kxm045} is a widely-used approach to estimate the sparse precision matrix using $\ell_1$ norm regularization under \ac{gmrf} models, where this precision represents a graph topology. \ac{lgmrf} models are a specific type of \ac{gmrf}, in which the precision is a Laplacian matrix. \ac{lgmrf} models have been widely studied 
\cite{Xiaowen2015Laplace,Egilmez_Pavez_Ortega_2017}. The formulation of the considered \ac{cmle} approach can be used for the estimation of complex-valued Laplacian matrices under the \ac{glasso} setting. In this case, the optimization problem is formulated similarly to \eqref{optimization_p4},  where the matrices $\Gmat$ and $\tilde{\Bmat}$ are the precision matrices 
in the negative log-likelihood $\Psimat(\Xmat,\Gmat,\tilde{\Bmat})$, as formulated in \cite{Tugnait_2019}. However, the proposed \textcolor{black}{\ac{admm}} algorithm has been developed for $\Psimat(\Xmat,\Gmat,\tilde{\Bmat})$ that can be expressed in the general quadratic form of \eqref{general_quadtratic_model}, which results in a tractable algorithm. This quadratic form excludes the \ac{glasso} formulation.
 An interesting issue for future research is the development of tractable algorithms for solving the CMLE optimization in \eqref{optimization_p4} with the \ac{lgmrf} model. It should be noted that the extension of existing \ac{gmrf} algorithms for \ac{lgmrf} is not trivial. For example, 
 in contrast with \ac{glasso}  algorithms, the $\ell_1$ norm penalty function in \ac{glasso} is ineffective in promoting sparsity under Laplacian constraints \cite{ying2020does,Yakov2023Laplace}. 

\subsubsection{Identifiability conditions and measurement avliability}
\label{Identifiability Conditions and Measurement Availability}
As can be seen from \eqref{G1 matrix} and \eqref{update equation for B_1 step - general}, a necessary condition to implement the \ac{admm} algorithm is that
the matrices $\Hmat_{1}+\rho\Emat$ and $\Hmat_{4}+\rho\Emat$ are full rank matrices. Due to the form of the matrix $\Emat$ defined in \eqref{Adef for G}, this matrix is a full-rank matrix (since $\Emat$ also includes in the sum the identity matrix $\Imat_{M^2}$). However, due to the $\ell_1$ norm regularization term in the objective function in \eqref{optimization_p4}, if there is not enough data, then the influence of the likelihood term in \eqref{optimization_p4}, $\psi(\Xmat,\Gmat,\tilde{\Bmat})$, would vanish, and the solution will converge to zero Laplacian matrices, which is non-informative. 
Without the sparsity assumption (and as a result, without the regularization term) the optimization problem that aims to minimize the linear quadratic model in \eqref{general_quadtratic_model} is identifiable
as long as the matrices $(\avec_{2}^T \otimes\Amat_{1})$ and $(\avec_{3}^T\otimes \Amat_{2})$ are full $M$ rank, and enough data is available. 

In our model, we assume the general form of the data, given by the matrix $\Xmat$, which determines the values of the matrices $(\avec_{2}^T \otimes\Amat_{1})$ and $(\avec_{3}^T\otimes \Amat_{2})$. The matrix $\Xmat$ can be collected from all the nodes or from a subset of nodes, where the number and spatial distribution of available measurements in the system are significant. 
In the following section,  we describe power system models with the availability of measurements from every bus in the system. 
 However, this is not a necessary condition for implementing the proposed ADMM.
Consideration of the sparsity condition for recovery, and sensor allocation to guarantee the identifiability of the system are interesting issues for further research in this area. 
\color{black}


\section{Implementation for Power System models}
\label{models_section}
In this section, we demonstrate the implementation of the  \textcolor{black}{\ac{admm}} algorithm for power flow models that are widely used in power system analysis and monitoring \cite{Abur_Gomez_book}. These models (the \ac{ac}, \ac{dlpf}, and \ac{dc} models in Subsections \ref{AC MODEL SECTION}, \ref{DLPF_subsection}, and 
\ref{DC_subsection}, respectively) are based on 
   the representation of an electric power system as a set of buses connected by transmission lines, and on the use of Kirchhoff’s and Ohm’s laws to determine the 
voltage (state) at each bus.  However, they differ in the level of detail and mathematical complexity. 
This problem can be interpreted as the inverse power flow problem, which infers a nodal admittance matrix from voltage and power phasors, and is of great importance \cite{Ye_Ardakanian_Low_Yuan_2016}. \textcolor{black}{In implementing the ADMM algorithm,  the convexity of $\psi(\Xmat,\Gmat,\tilde{\Bmat})$ is essential. 
For comprehensive insights into the proof of convexity and Hessian calculations related to strict convexity, see the supplemental materials. An interesting scenario not covered here is the case where measurements come from a variety of sensor types, such as \ac{pmu} and smart meters, and where measurements are not universally available across all nodes. In this paper, we assume that measurements are available from all nodes. Further research is needed on the use of partial observations and its effect on the topology identification accuracy.}

\subsection{\ac{ac} model}
\label{AC MODEL SECTION}
In this subsection, we consider the commonly-used nonlinear \ac{ac} power flow model \cite{Abur_Gomez_book,Giannakis_Wollenberg_2013}.
In this model, the power flow on each transmission line is determined by the complex power injected at each end of the line, the voltages, and the line's complex-valued admittance. 
Accordingly, the noisy measurements of 
power injections at the $M$ buses over $N$ time samples
can be written
  as follows \cite{Abur_Gomez_book}:
\be
\label{noisy_model}
\pvec[n]+j \qvec[n]={\text{diag}}(\vvec[n])(\Gmat+j\tilde{\Bmat})\vvec^{*}[n]
+ \etavec[n],
\ee
$n=0,\ldots,N-1$,
 where $n$ is the time index,
 $\{\pvec[n],\qvec[n]\}_{n=1}^N$ are the active and reactive power vectors, and
$\{\vvec[n]\}_{n=1}^N$
denotes the phasor voltages.
Here,  $p_m[n]+jq_m[n]$  and $v_m[n]$ are the complex-valued power and voltage at the $m$th bus measured at time $n$.
The noise sequence, $\{\etavec[n]\}_{n=0}^{N-1}$, is assumed to be an \ac{i.i.d} zero-mean complex circularly symmetric Gaussian noise with a known \textcolor{black}{positive definite} covariance matrix, $\Rmat_{\eta}$.

In this case,  the measurements $\Xmat$ consist of  $\{\pvec[n],\qvec[n],\vvec[n]\}_{n=1}^N$, that can be obtained, for instance, by \ac{pmu} \cite{PMUbook}.
Thus, 
the objective function 
 (i.e. {\textcolor{black}{negative}} log-likelihood of the model in \eqref{noisy_model}) for finding the 
\ac{cmle} of $\Gmat$ and $\tilde{\Bmat}$
is
\begin{eqnarray}
    \label{tttt}
        \psi_{\text{AC}}(\Xmat,\Gmat,\tilde{\Bmat})\hspace{6cm}\nonumber\\ \define \sum_{n=0}^{N-1}\| \pvec[n] +j\qvec[n]- {\text{diag}}(\vvec[n])(\Gmat+j\tilde{\Bmat})\vvec^{*}[n]\|_{\Rmat_{\eta}^{-1}}^2\nonumber \\\quad= \sum\nolimits_{n=0}^{N-1}\left\| \pvec[n] + j\qvec[n]\right.\hspace{4.25cm}\nonumber \\
-(\vvec^{H}[n] \otimes {\text{diag}}(\vvec[n]))  {\text{Vec}}(\Gmat+j\tilde{\Bmat})\|_{\Rmat_{\eta}^{-1}}^2,\hspace{1.25cm}
\end{eqnarray}
where the last equality is obtained by using the properties of the vec operator and the  Kronecker product. It can be seen that  \eqref{tttt} has the general structure described in \eqref{general_quadtratic_model}.
Therefore, for the \textcolor{black}{\ac{admm}} approach, the matrices and vectors  from \eqref{H1_H2_def}-\eqref{h1_h2_def} for \textbf{Stage 1} in Subsection \ref{two_steps_ALM_subsection} satisfy
\begin{eqnarray}
        \label{AC matrices for ALM}
        \Hmat_{1} +j\Hmat_{2} \hspace{6cm}\nonumber\\ = 2 \sum_{n=0}^{N-1} (\vvec[n] \vvec^H[n])\otimes \left(\text{diag}(\vvec^{*}[n])\Rmat_{\eta}^{-1}\text{diag}(\vvec[n])\right) 
\end{eqnarray}
\begin{equation}
        \label{AC vectors for ALM}
        -\hvec_{1}+j \hvec_{2}=\sum_{n=0}^{N-1}{\text{Vec}}\left(\text{diag}(\vvec[n])\Rmat_{\eta}^{-1} 
 (\pvec[n]-j\qvec[n])
 \vvec^{H}[n]\right), 
\end{equation}
where
$\Hmat_i$, $i=1,2,3,4$ are real-valued matrices, and $\hvec_i$, $i=1,2$  are real-valued vectors, and their values can be extracted from \eqref{AC matrices for ALM} and \eqref{AC vectors for ALM}.
In addition, $\Hmat_{4} = \Hmat_{1}$ and $\Hmat_{3} = \Hmat_{2}$.

\subsection{\ac{dlpf} model}
\label{DLPF_subsection}
The \ac{dlpf} model is an approximation of the \ac{ac} power flow model from Subsection \ref{AC MODEL SECTION} \cite{7782382}.
In the \ac{dlpf} model, the power flow on each transmission line is determined by the voltage amplitude and angle difference between the buses at each end of the transmission line, as well as the line's conductance and susceptance. 
 According to this model, the noisy power injection measurements 
 at the $M$ buses over $N$ time samples
can be written 
 as follows \cite{Zhang2021}:
\begin{equation}
\label{measurements model of the DLPF}
\displaystyle \begin{cases}
	\displaystyle \pvec[n]=\tilde{\Bmat}\btheta[n] + \Gmat|\vvec[n]|+ \text{Re}\left\{\etavec[n]\right\} \\
	\displaystyle \qvec[n]=-\Gmat\btheta[n] + \tilde{\Bmat}|\vvec[n]| + \text{Im}\left\{\etavec[n]\right\}
\end{cases},
\end{equation}
$n=0,\dots,N-1$, where the real and imaginary parts of $\etavec$ are \ac{i.i.d} white Gaussian sequences with zero mean and covariance matrices $\frac{1}{2}\Rmat_{\eta}$, and $\btheta[n]$ is the angle of the phasor voltages $\vvec[n]$.
Thus, in this case we assume that the measurements $\Xmat$ consist of data of $\{\pvec[n],\qvec[n],|\vvec[n]|,\thetavec[n]\}_{n=1}^N$, which can be obtained, for instance, from \ac{pmu}.
Thus, 
the objective function $\psi(\Xmat,\Gmat,\tilde{\Bmat})$
 (i.e. {\textcolor{black}{negative}} log-likelihood function of \eqref{measurements model of the DLPF}) for finding the 
\ac{cmle} of $\Gmat$ and $\tilde{\Bmat}$
  for this case is
\begin{eqnarray}
\label{Objective DLPF model}
  \psi_{\text{DLPF}}(\Xmat,\Gmat,\tilde{\Bmat})\hspace{5.25cm}\nonumber\\=\sum_{n=0}^{N-1}\left\|
    \begin{bmatrix}
    \pvec[n]\\\qvec[n] 
    \end{bmatrix} - 
    \begin{bmatrix}
    \tilde{\Bmat}&\Gmat\\-\Gmat&\tilde{\Bmat}
    \end{bmatrix}
    \begin{bmatrix}
    \thetavec[n] \\ |\vvec[n]|
    \end{bmatrix}
    \right\|_{2\Imat_{2}\otimes \Rmat_{\eta}^{-1}}^{2} 
    .
\end{eqnarray}
Similar to \eqref{tttt},
 the measurement equation in \eqref{Objective DLPF model} can be written
with the general structure described in \eqref{general_quadtratic_model}.
Thus,
for the \textcolor{black}{\ac{admm}} approach, 
the matrices and vectors  from \eqref{H1_H2_def}-\eqref{h1_h2_def} for \textbf{Stage 1} in Subsection \ref{two_steps_ALM_subsection} are
\begin{equation}
    \begin{aligned}
        \label{DLPF matrices for ALM - part 1}
        &\Hmat_{1}=\scalebox{1}{$\displaystyle  2\sum_{n=0}^{N-1}\bigg(|\vvec[n]||\vvec^{T}[n]| + \thetavec[n]\thetavec^{T}[n]\bigg)\otimes\Rmat_{\eta}^{-1}$},
    \end{aligned}
\end{equation}
\begin{equation}
    \begin{aligned}
        \label{DLPF matrices for ALM - part 2}
        &\Hmat_{2}=\scalebox{1}{$\displaystyle    2\sum_{n=0}^{N-1}\bigg(\thetavec[n]|\vvec^{T}[n]|-|\vvec[n]|\thetavec^{T}[n]\bigg)\otimes\Rmat_{\eta}^{-1}$},
    \end{aligned}
\end{equation}
\begin{equation}
    \begin{aligned}
        \label{DLPF matrices for ALM - part 3}
        &\hvec_{1}=\scalebox{1}{$\displaystyle 2\sum_{n=0}^{N-1}\text{Vec}\bigg(\Rmat_{\eta}^{-1}(\qvec[n]\thetavec^{T}[n]-\pvec[n] |\vvec^{T}[n]|)\bigg)$},
    \end{aligned}
\end{equation}
\begin{equation}
    \begin{aligned}
        \label{DLPF matrices for ALM - part 4}
        &\hvec_{2}=\scalebox{1}{$\displaystyle -2\sum_{n=0}^{N-1}\text{Vec}\bigg(\Rmat_{\eta}^{-1}(\pvec[n]\thetavec^{T}[n]+\qvec[n]|\vvec^{T}[n]|)\bigg)$}.
    \end{aligned}
\end{equation}
In addition, $\Hmat_{4} = \Hmat_{1}$ and $\Hmat_{3} = -\Hmat_{2}$.
\subsection{\ac{dc} model}
\label{DC_subsection}
The \ac{dc} model \cite{Donmez2016Abur} is a commonly-used linearized model of the power flow equations. 
The noisy \ac{dc} power flow model is given by \cite{Abur_Gomez_book,Giannakis_Wollenberg_2013}
\be
\label{DC_model}
\pvec[n]=\tilde{\Bmat}\btheta[n] + \text{Re}\left\{\etavec[n]\right\},
\ee
$n=0,\dots,N-1$, where  $\text{Re}\left\{\etavec[n]\right\}$ is an \ac{i.i.d} white Gaussian sequence with zero mean and covariance matrix $\frac{1}{2}\Rmat_{\eta}$, and $\btheta[n]$ is the angle of  $\vvec[n]$. 
Thus, in this case we assume that the measurements $\Xmat$ consist of data of $\{\pvec[n],\thetavec[n]\}_{n=1}^N$.
Hence, 
the objective function $\psi(\Xmat,\tilde{\Bmat})$
 (i.e. {\textcolor{black}{negative}} log-likelihood function of \eqref{measurements model of the DLPF}) for finding the
\ac{cmle} of $\tilde{\Bmat}$
 is
\begin{eqnarray}
\label{Objective DC model}
    \psi_{\text{DC}}(\Xmat,\tilde{\Bmat})
    =\sum\nolimits_{n=0}^{N-1}\|\pvec[n]-\tilde{\Bmat}\thetavec[n]\|_{2\Rmat_{\eta}^{-1}}^{2}
    \hspace{2cm}\nonumber\\
    =\sum\nolimits_{n=0}^{N-1}\|\pvec[n]-
     \left(\thetavec^T[n] \otimes \Imat_M\right)  {\text{Vec}}(\tilde{\Bmat})\|_{2\Rmat_{\eta}^{-1}}^{2},
\end{eqnarray}
where the last equality is obtained by using the properties of the vec operator and the  Kronecker product.

It can be seen that the measurement equation in \eqref{Objective DC model} has the general structure described in \eqref{general_quadtratic_model}.
Based on the \ac{dc} model, we are only capable of estimating the imaginary part of the admittance matrix from \eqref{Y_def},   $\tilde{\Bmat}$, and cannot estimate $\Gmat$. Thus,
for the \textcolor{black}{\ac{admm}} approach, 
the relevant matrices and vectors  from \eqref{H1_H2_def}-\eqref{h1_h2_def} for \textbf{Stage 1} in Subsection \ref{two_steps_ALM_subsection} are
\begin{equation}
\label{DC matrices for ALM}
	\displaystyle \Hmat_{3}=\zerovec, {\text{ and }}\Hmat_{4} = 2\sum\nolimits_{n=0}^{N-1}\thetavec[n]\thetavec^{T}[n]\otimes\Rmat_{\eta}^{-1} 
 \end{equation}
 \begin{equation}
 \label{DC vectors for ALM}
\hvec_{2} = -2\sum\nolimits_{n=0}^{N-1}\text{Vec}(\Rmat_{\eta}^{-1}\pvec[n]\thetavec^{T}[n]).
\end{equation}

In \cite{Grotas_Routtenberg_2019}, the \ac{dc} measurement model has been used to jointly estimate the states ($\thetavec[n]$) and the real-valued susceptance matrix ($\Bmat$). It can be shown that the Lagrangian-based method suggested in \cite{Grotas_Routtenberg_2019} for estimating $\Bmat$, assuming the states are given as input, is a special case of the proposed  \textcolor{black}{\ac{admm}} algorithm for the DC model. 
In addition, here we provide a closed-form expression for the first step of the \textcolor{black}{\ac{admm}} approach in \eqref{update equation for B_1 step - general}, whereas in  \cite{Grotas_Routtenberg_2019}, a general, implicit gradient descent update equation is presented.  

\subsection{Discussion on power network characteristics}
\textcolor{black}{It is important to highlight the differences between the \ac{dn} and the \ac{tn}. TNs are generally extensively interconnected, requiring analysis as a cohesive whole. In contrast, \acp{dn} consist of many independent substations, each supplying power to several radial feeders, allowing for analysis as multiple independent units. Furthermore, \acp{tn} often have a high percentage of devices equipped with measurement capabilities, rendering them mathematically observable \cite{Abur_Gomez_book}, while DNs feature more load points than measurements\cite{7779155} and are often significantly less observable \cite{Giannakis_Wollenberg_2013}. Additionally, while the typical network size for \acp{tn} ranges from a few hundred to a couple of thousand buses, \acp{dn} can range from 10,000 to 100,000 electrical nodes, highlighting a substantial difference in scale. Thus, to apply the proposed methods in \acp{dn}, additional work is needed to address issues such as missing measurements and to integrate structural characteristics, including the radial network \cite{DekaDist2024}. Moreover, innovative approaches for exact topology and parameter estimation with minimal observability are also critical to enhancing DNs' operational efficiency and real-time market participation \cite{Park_Deka_Chertkov2018}. Additionally, methods that estimate the admittance matrix using linear regression based on active/reactive power injection and voltage magnitude data further contribute to addressing the challenges of open-loop network analysis with limited data \cite{Zhang2021}.}


 \section{Simulations}
 \label{simulations}

In this section, we illustrate the performance of recovering the admittance matrix using the algorithm developed in Section \ref{Estimation Methods for General Objective} for all the models described in Section \ref{models_section}.  We implement the proposed methods for the IEEE 33-bus system,  with parameters from \cite{iEEEdata}. The data for the phasor voltages are also generated using Matpower \cite{zimmerman2011matpower}, a software package for power system analysis. 
{\textcolor{black}{We first numerically evaluate the joint sparsity of the real and imaginary parts of the admittance matrix in typical power systems in Subsection \ref{validation_subsec}. Then, in Subsection \ref{setting_subsec}, we describe the different methods for comparison and the setting of the simulations. }}
We show the results of two scenarios, where the measurements were generated using 1) the \ac{dlpf} model (Subsection \ref{scenario2_subsec}); and 2) the \ac{ac} model (Subsection \ref{AC_mes_subsec}). {\textcolor{black}{Finally, the run-time performance of the different algorithms is presented in Subsection \ref{runtime_subsec}.}}
\vspace{-0.25cm}
\subsection{Experimental validation: the joint sparsity of the admittance matrix in typical networks}
 \label{validation_subsec}
In order to experimentally validate the sparsity patterns of $\tilde{\Bmat}$ and $\Gmat$ in typical electrical networks, 
in Table \ref{tabSmooth} we present the size of the support of the two Laplacian matrices (i.e. the connectivity pattern of the graph based on $\Gmat$ and $\tilde{\Bmat}$)
for the IEEE test cases \cite{iEEEdata}.
In addition, we compute the F-score between the two matrices \cite{sokolova} (which measures the differences in the sparsity patterns of $\tilde\Bmat$ and $\Gmat$), by treating 
one of the matrices ($\tilde\Bmat$ or $\Gmat$)  as the ground truth and the other as the estimate. 
This F-score takes values between $0$ and $1$, where, in this case, $1 $ indicates a situation where the supports of $\tilde{\Bmat}$ and of $\Gmat$ are identical.
It can be seen from this table that the matrices are almost (but not perfectly) jointly sparse.
 This assumption is more restrictive than assuming that each matrix is sparse with its own independent support. 
From the evaluation of the ratio between the matrix elements, it can also be seen that the magnitudes of the elements in $\Bmat$ are larger than the magnitudes of the elements in $\Gmat$.
Finally, in Fig. \ref{Fig1:B}, we illustrate the relations between the real and imaginary parts of the admittance matrix, $\Gmat$ and $\tilde{\Bmat}$, of the IEEE 14-bus system. This figure shows that the supports of the two matrices are sparse, with similar but not identical patterns. 
\vspace{-0.25cm}
\begin{table}[hbt]
\begin{center}
\begin{tabular}{|c|c|c|c|c|c|}
\hline
\text{}&\multicolumn{5}{|c|}{\textbf{IEEE test case system}} \\
\cline{2-6} 
\textbf{Measures} & \textbf{\textit{14-bus}}& \textbf{\textit{33-bus}}& \textbf{\textit{57-bus}}& \textbf{\textit{118-bus}}& \textbf{\textit{145-bus}} \\
\hline
    $|\xi_\Gmat|$ &15 &32 &62 &170 &409 \\
\hline
$|\xi_\Bmat|$ &20 &32 &78 &179 &422\\
\hline
$\textbf{F-score}$ &0.875 & 1 &0.9160 & 0.9807 & 0.9836  \\
\hline
\scalebox{0.8}{$\|\text{Vec}_{\ell}(\Gmat),\text{Vec}_{\ell}(\tilde{\Bmat})\|_{p,0}$} & 20 & 32& 78& 179& 422 \\
\hline
\end{tabular}
\caption{Properties of IEEE test case systems: the first and second rows display the number of edges (nonzeros in $\Gmat$ and $\Bmat$). The third row is the F-score between the matrices,
which measures the differences in their sparsity patterns.
\textcolor{black}{Finally, the fourth row is the row sparsity  of $\|\text{Vec}_{\ell}(\Gmat),\text{Vec}_{\ell}(\tilde{\Bmat})\|_{p,0}$.} }
\label{tabSmooth}
\end{center}
\end{table}

\begin{figure}[hbt]
    \centering
    \begin{subfigure}[b]{0.21\textwidth}
        \centering
        \includegraphics[width=\textwidth]{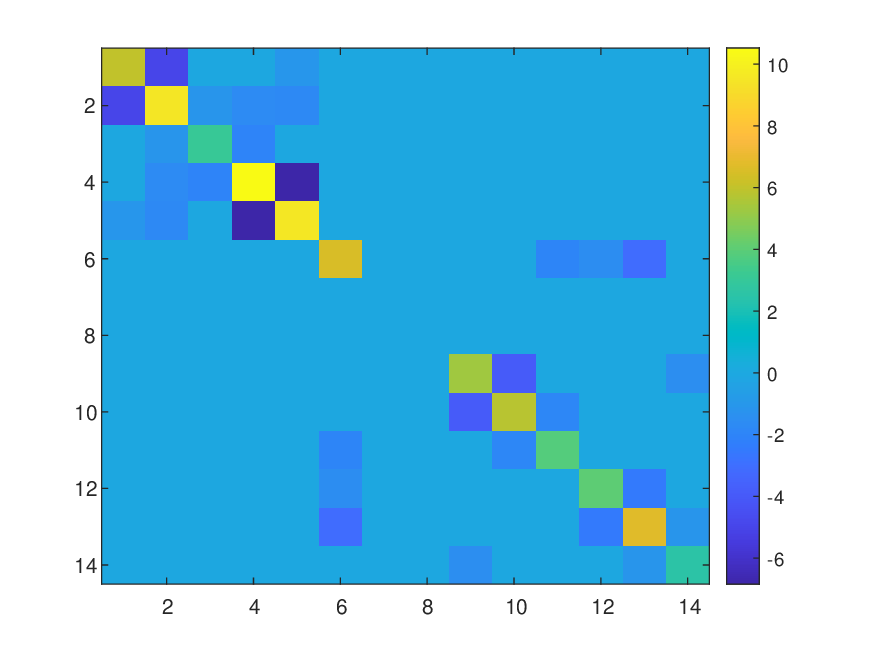}
        \subcaption{\label{Fig1:G}}
    \end{subfigure}
    \hfill
    \begin{subfigure}[b]{0.21\textwidth}
        \centering
        \includegraphics[width=\textwidth]{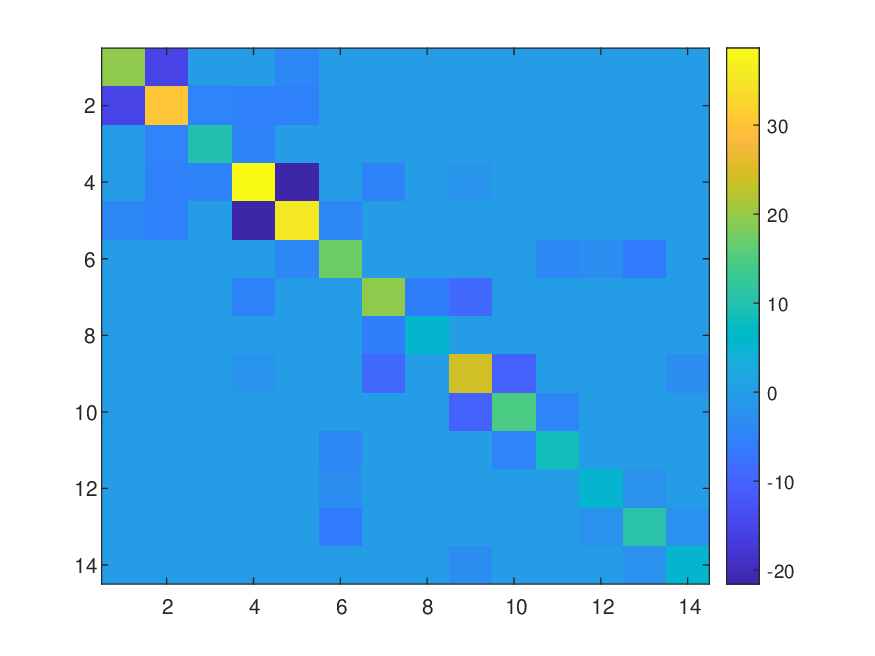}
        \subcaption{\label{Fig1:B}}
    \end{subfigure}
    \caption{Illustration of the real (a) and imaginary (b) parts of the admittance matrix for IEEE 14-bus system.}
    \label{FigGB14BUS}
\end{figure}

\color{black}
\vspace{-1cm}
\subsection{\textcolor{black}{Setting}}
\label{setting_subsec}
We compare the performance of the following methods:\renewcommand{\theenumi}{\arabic{enumi}}
\begin{enumerate}
\item the block coordinate descent (BCD) method in \cite{Li_Poor_Scaglione_2013} with the regularizer $\mu = 14$ and the step size $\beta = 10^{-4}$,
\item the ADMM algorithm for the \ac{dc} model in \cite{Anwar_Mahmood_Pickering_2016} with the penalty $\rho = 0.0001$ and the regularizer $\lambda = 5$,
\item the \ac{tls} algorithm for the DLPF model in \cite{Zhang2021}.
\item the proposed \textcolor{black}{\ac{admm}} method - implemented by Algorithm \ref{ALGO2} with the parameters:
the maximal number of iterations $N_{max}=100$, the convergence parameter $\epsilon=0.01$,  
and the penalty parameter $\rho=0.0001$, which was tuned experimentally.
The implementations of the 
 \textcolor{black}{\ac{admm}} method for the \ac{ac}, \ac{dlpf}, and \ac{dc} models (from Subsections \ref{AC MODEL SECTION}, \ref{DLPF_subsection}, and
\ref{DC_subsection}) are denoted by 
\textcolor{black}{\ac{admm}}-AC, \textcolor{black}{\ac{admm}}-DLPF, and \textcolor{black}{\ac{admm}}-DC, respectively.
\end{enumerate}
The first two methods jointly estimate $\Bmat$ and the angle of the phasor voltages $\{\vvec[n]\}_{n=0}^{N-1}$,  $\{\thetavec[n]\}_{n=0}^{N-1}$, based on the \ac{dc} model with measurements of the real power injection data at each bus ($\{\pvec[n]\}_{n=0}^{N-1}$). Thus, in order to have a fair comparison, we insert $\vvec[n]$ as an input to those methods. However, these methods are incapable of estimating $\Gmat$.
The third method is capable of estimating both $\Gmat$ and $\Bmat$, and also the angle of $\vvec[n]$ based on the DLPF model with measurements of the complex power injection data at each bus ($\{\pvec[n],\qvec[n]\}_{n=0}^{N-1}$). 
In  methods $1,2,4,5$, we initialized the matrices $\Gmat$ and $\tilde{\Bmat}$ using our method from Algorithm \ref{ALGO3}.

The performance is compared in terms of the MSE  of the matrix estimation, defined as
$\frac{1}{M^2}\text{Trace}((\hat{\Amat}-\Amat)^{T}(\hat{\Amat}-\Amat))$, 
where $\Amat$ is the matrix that is estimated and $\hat{\Amat}$ is the estimator, and in terms of
the F-score.
The F-score
measures the error probability in the connectivity of the matrices $\Bmat$ and $\Gmat$ that represent the sparsity pattern, and is given by \cite{Egilmez_Pavez_Ortega_2017}
\be
\label{FS}
\textrm{FS}= \frac{2\textrm{tp}}{2\textrm{tp}+\textrm{fp} +\textrm{fn}},
\ee
which is based on true-positive (tp), false-positive (fp), and false-negative (fn) probabilities of detection of graph edges in the estimated Laplacian compared to the ground truth support of the Laplacian matrix. 
This F-score takes values between $0$ and $1$, where $1 $ indicates perfect classification, i.e. all the connections and disconnections between the nodes in the underlying ground-truth graphs are correctly identified.
We present the \ac{mse} and F-score of the different estimators of $\Gmat$ and $\Bmat$ as a function of the \ac{snr}, defined as
${\text{SNR}} = 10\log(\frac{1}{MN \sigma^{2}} \sum_{n=0}^{N-1}\left| \pvec[n]+j\qvec[n]\right|^2)$, where $N=800$ in all simulations.  All simulations are averaged over 100 Monte-Carlo simulations for each scenario.

\subsection{Scenario 1: \ac{dlpf} measurement model}
\label{scenario2_subsec}
In this scenario, the power injection data was generated based on the \ac{dlpf} model from \eqref{measurements model of the DLPF}. Although this model is less accurate than the AC model in the next subsection, it is shown here in order to have a fair comparison with the \ac{tls} method that was developed based on this model.
Figures \ref{Fig2:a}-\ref{Fig2:b} show the MSE and F-score of the estimators of the matrix $\Gmat$ versus SNR. Since the \ac{dc}-based methods (i.e. BCD-DC and ADMM-DC) are only capable of estimating $\Bmat$, they are not included in these results. Figures \ref{Fig3:a}-\ref{Fig3:b} show the MSE and F-score of the estimators of the matrix $\Bmat$ versus SNR.

Figures \ref{Fig2} and \ref{Fig3} show that the \textcolor{black}{\ac{admm}}-DLPF method from Subsection \ref{DLPF_subsection} outperforms other examined methods in terms of MSE. In this tested scenario, the \textcolor{black}{\ac{admm}}-DLPF has better performance than the \textcolor{black}{\ac{admm}}-AC method since the measurement model has been generated according to the \ac{dlpf} model, which is less accurate and is only tested here in order to have a fair comparison with the \ac{tls} method.
As a result,  it can be seen in Figs. \ref{Fig2:a} and \ref{Fig3:a} that the methods that have been developed for the \ac{dlpf} model, i.e. the \ac{tls} and the proposed \textcolor{black}{\ac{admm}}-DLPF method, 
 show a monotonic decrease in \ac{mse} as the \ac{snr} increases, while the other methods converge to an almost constant value for SNR higher than $15$dB.
The \ac{tls} method achieves good performance both in F-score and MSE for high \ac{snr} values (SNR$>35$dB), is significantly faster than the other examined methods, and avoids the optimization process and the need to set tuning parameters. However, it does not converge and
has poor performance for lower \acp{snr}.


\subsection{Scenario 2: AC measurement model}
\label{AC_mes_subsec}
In this scenario, the power injection data was generated based on the \ac{ac} model from \eqref{noisy_model}.
Figures \ref{Fig4:a}-\ref{Fig4:b} show the MSE and F-score of the estimators of the matrix $\Gmat$ versus SNR. 
Since the \ac{dc}-based methods (i.e. BCD-DC and ADMM-DC) are only capable of estimating $\Bmat$, they are not included in these results. Figures \ref{Fig5:a}-\ref{Fig5:b} show the  MSE and F-score of the estimators of$\Bmat$ versus SNR.
The \ac{tls} method 
was not applicable in this scenario due to resolution limitations.

Figures \ref{Fig5:a}-\ref{Fig5:b} demonstrate that the \textcolor{black}{\ac{admm}}-AC algorithm from Subsection \ref{AC MODEL SECTION} outperforms the DC-based methods (BCD-DC, ADMM-DC, \textcolor{black}{\ac{admm}}-DC) and the \textcolor{black}{\ac{admm}}-DLPF method
in terms of both MSE and F-score for estimating $\Bmat$. This method was developed under the approximation of the linearized DC model or the approximation of the DLPF model. Thus, these results demonstrate the value of using a more accurate description of the power flow model,  i.e. the AC model, versus the linear models. Moreover, the proposed \textcolor{black}{\ac{admm}}-AC, and \textcolor{black}{\ac{admm}}-DLPF algorithms recover the full complex-valued admittance, which provides a more accurate description of the topology. 
\begin{figure}[hbt]
    \centering
	\subcaptionbox{\label{Fig2:a}}[\linewidth]
	{ \includegraphics[width=0.85\linewidth]{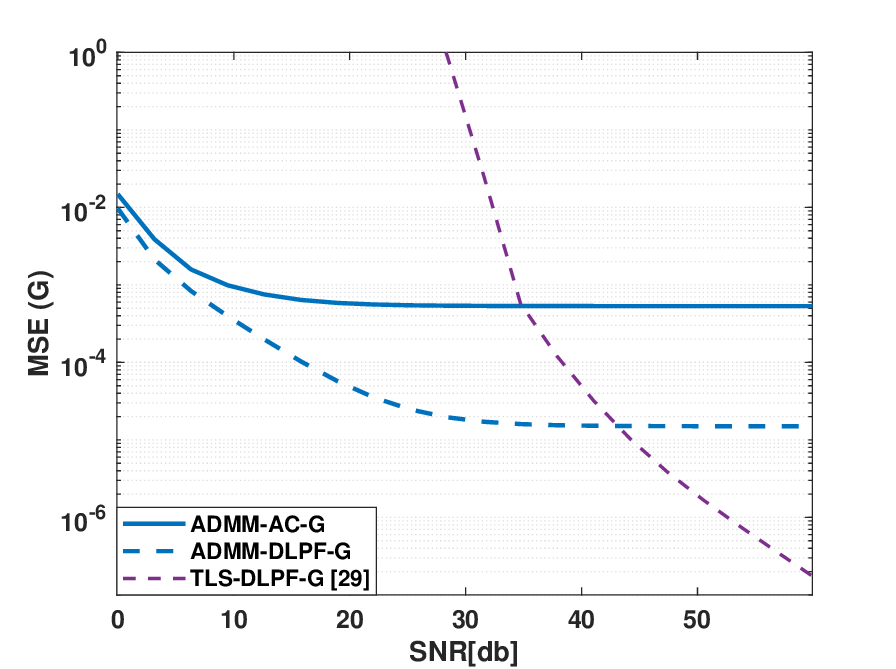}	}
      \subcaptionbox{\label{Fig2:b}}[\linewidth]
	{ \includegraphics[width=0.85\linewidth]{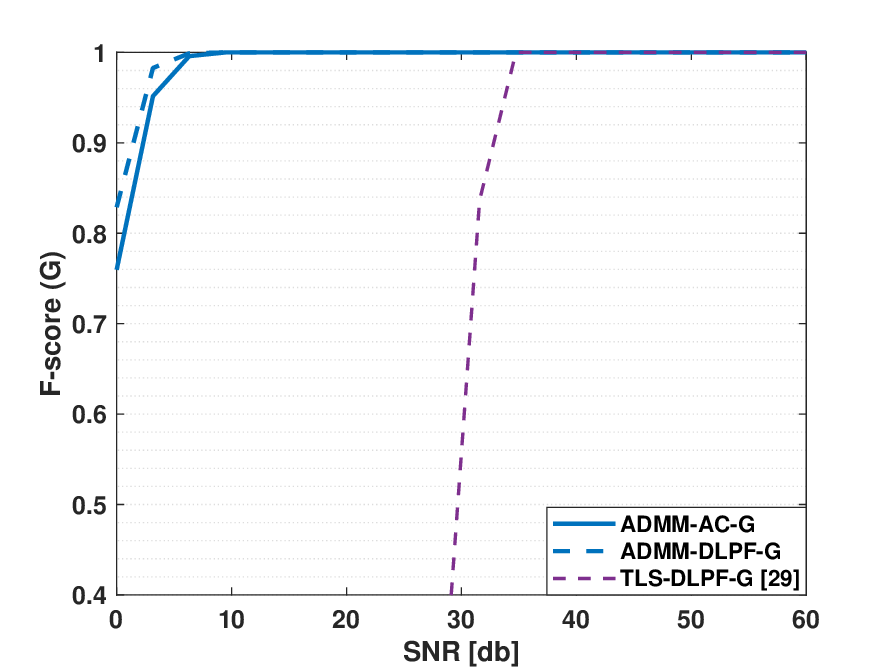}	}
	\caption{The MSE (a) and F-score (b) of the different estimators of  $\Gmat$ for IEEE 33-bus system, based on the \ac{dlpf} measurement model.}
	\label{Fig2}
\end{figure}
\begin{figure}[hbt]
    \centering
	\subcaptionbox{\label{Fig3:a}}[\linewidth]
	{\includegraphics[width=0.85\linewidth]{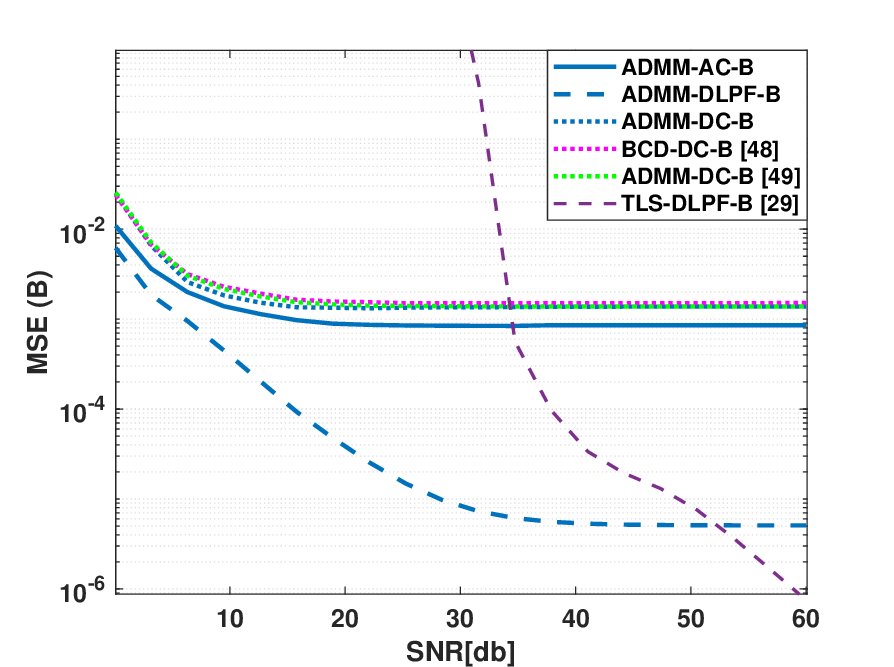}	}
        \subcaptionbox{\label{Fig3:b}}[\linewidth]
	{\includegraphics[width=0.85\linewidth]{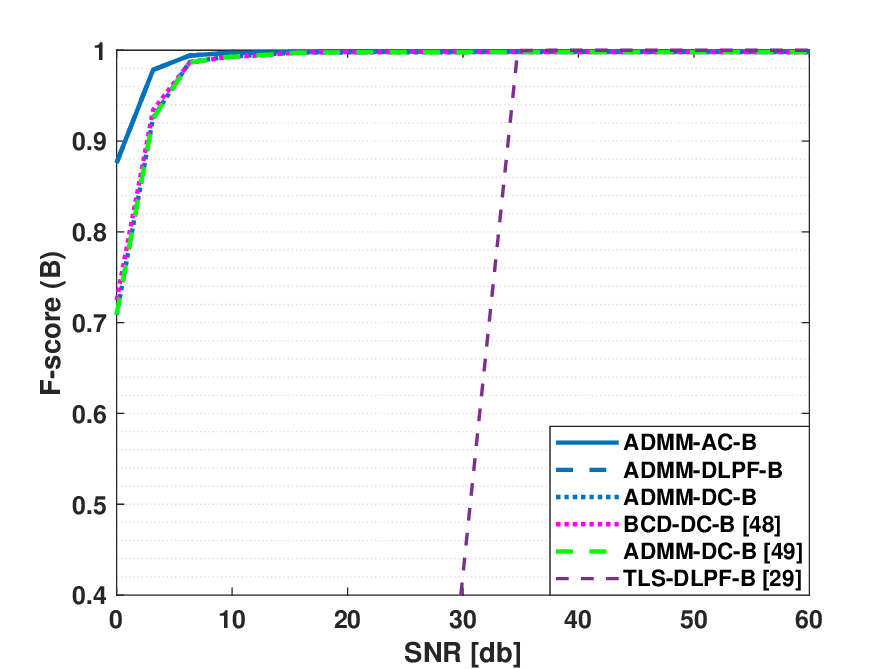}	}
	\caption{The MSE (a) and F-score (b) of the different estimators of  $\Bmat$ for IEEE 33-bus system, based on the \ac{dlpf} measurement model.}
	\label{Fig3}
\end{figure}




\begin{figure}[hbt]
    \centering
	\subcaptionbox{\label{Fig4:a}}[\linewidth]
	{\includegraphics[width=0.85\linewidth]{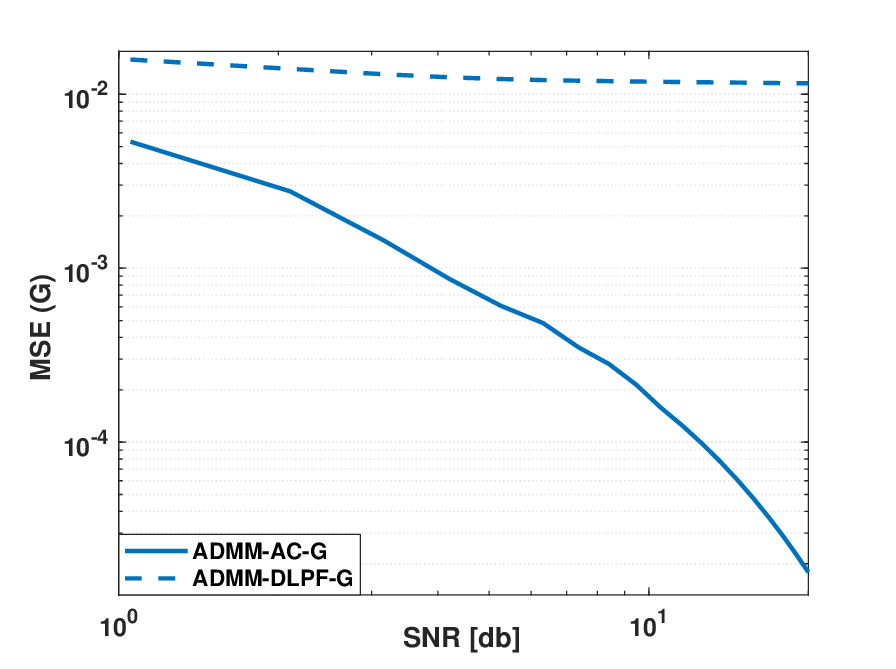}	}
        \subcaptionbox{\label{Fig4:b}}[\linewidth]
	{\includegraphics[width=0.85\linewidth]{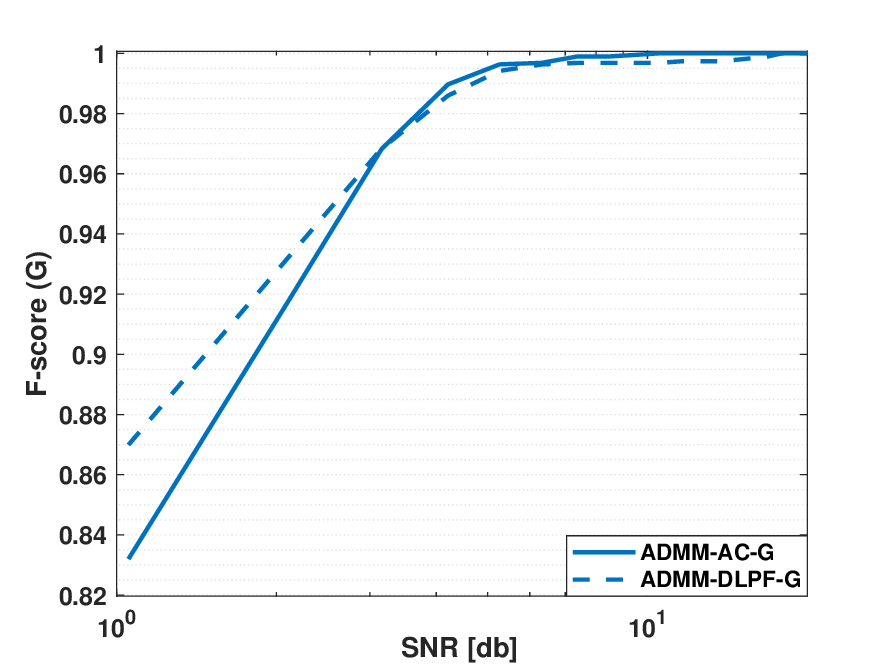}	}
	\caption{The MSE (a) and F-score (b) of the different estimators of  $\Gmat$ for IEEE 33-bus system, based on the \ac{ac}  model.}
	\label{Fig4}
\end{figure}

\begin{figure}[hbt]
    \centering
	\subcaptionbox{\label{Fig5:a}}[\linewidth]
	{\includegraphics[width=0.85\linewidth]{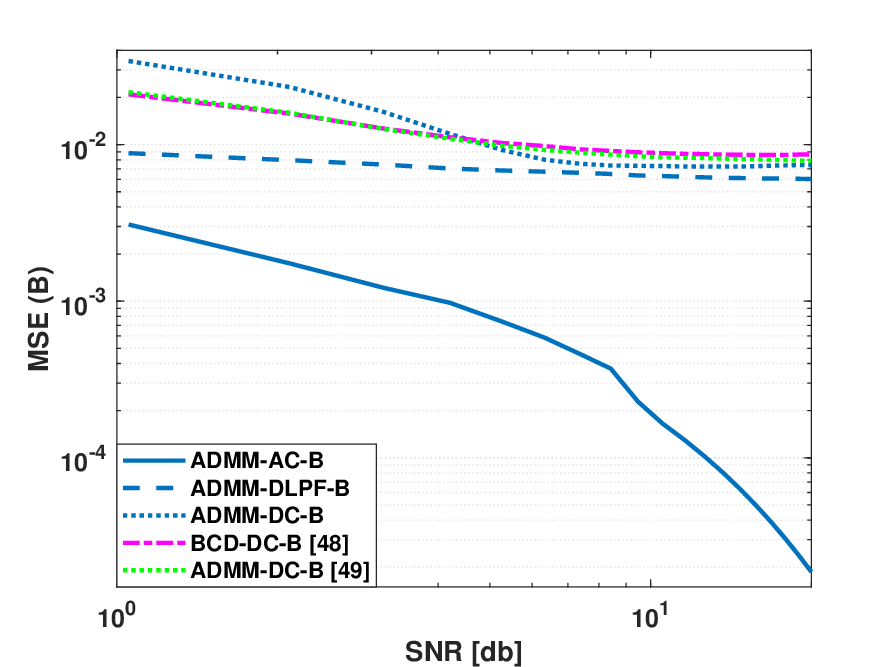}	}
        \subcaptionbox{\label{Fig5:b}}[\linewidth]
	{\includegraphics[width=0.85\linewidth]{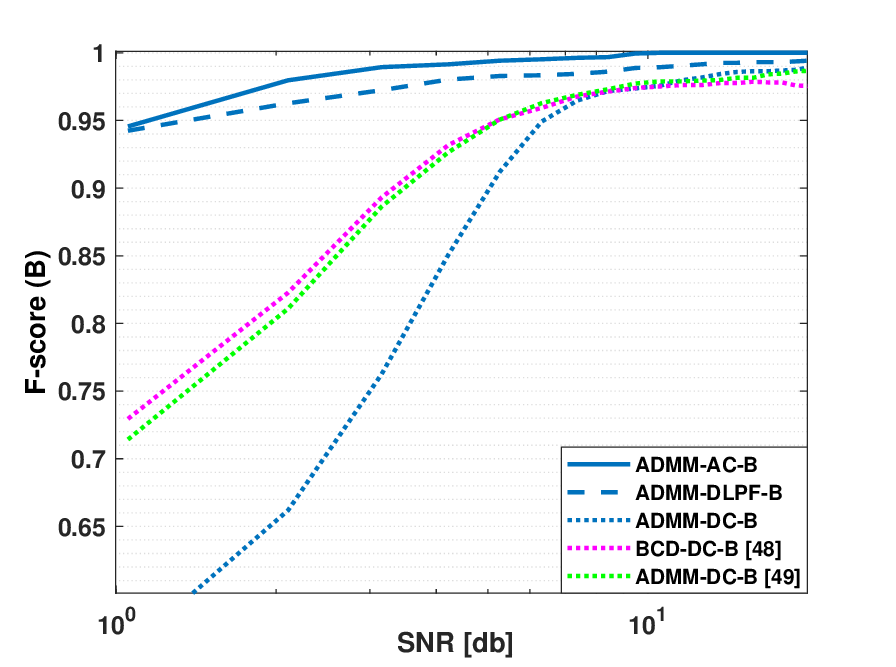}	}
	\caption{The MSE (a) and F-score (b) of the different estimators of  $\Bmat$ for IEEE 33-bus system, based on the \ac{ac} model.}
	\label{Fig5}
\end{figure}

\color{black}
\subsection{Run-time}
\label{runtime_subsec}
The computational complexity of the different methods is examined by the run-time required to execute the algorithm of the method. For our simulations, we utilized an Intel(R) Core(TM) i7-5820K CPU @ 3.30 GHz processor with 64.0 GB of installed RAM, running on a Windows 10 Enterprise Edition system (22H2 version). 
In Table \ref{table:algorithm_run_times}, we present the run-time required for different topology identification methods and IEEE test cases using data generated from the \ac{ac} model. 
It can be seen that the run-time of the methods increases as the system size increases.
Additionally,
the scalability of the ADMM-based methods, 
is reflected in the run-time performance presented in Table \ref{table:algorithm_run_times}. The ADMM-DLPF and ADMM-DC, in particular, maintain competitive run times as the system size grows, which is essential for their applicability in large-scale power systems. 

\begin{table}[hbt]
\begin{center}
\begin{tabular}{|l|c|c|c|c|c|}
\hline
\text{}&\multicolumn{4}{|c|}{\textbf{IEEE test case system}} \\
\cline{2-5} 
\textbf{Algorithm} & \textbf{\textit{14-bus}} & \textbf{\textit{33-bus}} & \textbf{\textit{57-bus}} & \textbf{\textit{118-bus}} \\
\hline
ADMM-AC  & 0.0145& 0.0362 & 0.1395 & 2.0273 \\
\hline
ADMM-DLPF  & 0.0122 & 0.0150 & 0.0313 & 0.2314 \\
\hline
ADMM-DC  & 0.0034 & 0.0035 & 0.0079 & 0.0902 \\
\hline
ADMM-DC \cite{Anwar_Mahmood_Pickering_2016}  & 0.0060 & 0.0315 & 0.3864 & 13.5056 \\
\hline
BSS-DC \cite{Li_Poor_Scaglione_2013}  & 0.0021& 0.0025 & 0.0028 & 0.0063 \\
\hline
TLS \cite{Zhang2021}  & 0.0165 & 0.0212 & 0.0298 & 0.1062 \\
\hline
\end{tabular}
\end{center}
\vspace{-0.25cm}
\caption{\color{black}Run-times of the algorithms for different IEEE test cases in [sec]}
\label{table:algorithm_run_times}
\vspace{-0.25cm}
\end{table}

\color{black}

\section{Conclusions}
\label{conclusions}
In this paper, we have proposed a method for the estimation of complex-valued Laplacian matrices with a joint sparsity pattern. 
We formulated the CMLE for this estimation problem as the solution of a convex regularized and constrained optimization problem,  {\textcolor{black}{with a group sparsity constraint. By utilizing the mixed $\ell_{2,1}$ norm relaxation, we
 developed the associated ADMM algorithm to solve this problem}}. We analyzed the computational complexity of the algorithm and introduced a sample-covariance-based initialization approach. We applied the proposed ADMM algorithm to the problem of estimating the admittance matrix in a power system in three commonly-used observation models of the power flow equations. Numerical simulations, using the IEEE 33-bus model test case, show that the proposed {\textcolor{black}{ADMM algorithm}} successfully reconstruct{\textcolor{black}{s}} the topology (i.e. the admittance matrix) and outperform{\textcolor{black}{s}} existing approaches in terms of MSE and F-score.  The proposed initialization method also improves the performance of existing methods developed for the DC model, BCD method \cite{Li_Poor_Scaglione_2013}  and ADMM algorithm \cite{Anwar_Mahmood_Pickering_2016}. 
Issues for further work in this area include extending the proposed methods to handle blind scenarios, where some voltage and/or power measurements are missing and need to be estimated as well {\textcolor{black}{(see, e.g. \cite{dinesh2023complex})}}. Additionally, it is crucial to develop performance bounds for the estimation approach. Furthermore, the development of \ac{gsp} tools to handle the complex-valued Laplacian matrix is expected to be useful for extensive applications in signal processing. \textcolor{black}{Additionally, future work is warranted to deal with the practical aspects of implementing the proposed method 
on DNs, which are often characterized by 
unique operational and structural challenges, as well as the integration of diverse sensor types to
enhance the applicability of our method to a broader range of practical scenarios in power systems.}

\vspace{-0.25cm}

\bibliographystyle{IEEEbib}

\newpage
\begin{center}
{\Large{Supplemental Material for the paper
Estimation of Complex-Valued Laplacian Matrices for Topology Identification in Power Systems}}
\author{Morad Halihal, Tirza Routtenberg, \IEEEmembership{Senior Member, IEEE}, and  H. Vincent Poor \IEEEmembership{Fellow Member, IEEE}\vspace{-0.25cm}}

\end{center}

In this report, we present detailed mathematical proofs that support the theoretical underpinnings of our work on the estimation of complex-valued Laplacian matrices for power systems in \cite{Morad_2024}. All the definitions and notations are presented in \cite{Morad_2024}. Specifically, we provide a rigorous proof of the convexity (not strictly) of a proposed objective function. Furthermore, we delve into the derivation of the Hessian matrix to affirm the convex nature of our problem formally. A part of our supplementary material is the proof of positive definiteness of the matrix $\Emat$, ensuring the stability and reliability of our estimation process.

\section{Proof of Convexity (not strict)}
\label{A1: Convexity Proof}
We aim to show that for all $\alpha \in [0,1]$, and for all $\Gmat_{1},~\Gmat_{2},~\tilde{\Bmat}_{1},~\tilde{\Bmat}_{2}\in\mathbb{R}^{M\times M}$:
\begin{equation}
    \begin{aligned}
    &\psi(\Xmat;\alpha\Gmat_{1} + (1-\alpha)\Gmat_{2},\alpha\tilde{\Bmat}_{1} + (1-\alpha)\tilde{\Bmat}_{2}) \\&\quad\leq  \alpha\psi(\Xmat;\Gmat_{1},\tilde{\Bmat}_{1}) + (1-\alpha)\psi(\Xmat;\Gmat_{2},\tilde{\Bmat}_{2}),        
    \end{aligned}
\end{equation}
\begin{equation}
    \begin{aligned}
    &\psi(\Xmat;\alpha\Gmat_{1} + (1-\alpha)\Gmat_{2},\alpha\tilde{\Bmat}_{1} + (1-\alpha)\tilde{\Bmat}_{2})\\&\quad =  \frac{1}{2}\|\avec_{1}+(\avec_{2}^{T}\otimes \Amat_{1})\text{Vec}(\alpha\Gmat_{1} + (1-\alpha)\Gmat_{2}) \\&\quad +(\avec_{3}^{T}\otimes \Amat_{2})\text{Vec}(\alpha\tilde{\Bmat}_{1}+(1-\alphavec)\tilde{\Bmat}_{2})\|_{2}^{2} \\&\quad =\frac{1}{2}\|\alpha(\avec_{1}+\tilde{\Amat}_{1}\text{Vec}(\Gmat_{1})+\tilde{\Amat}_{2}\text{Vec}(\tilde{\Bmat}_{1}))\\&\quad+(1-\alpha)(\avec_{1}+\tilde{\Amat}_{1}\text{Vec}(\Gmat_{2})+\tilde{\Amat}_{2}\text{Vec}(\tilde{\Bmat}_{2}))\|_{2}^{2}.        
    \end{aligned}
\end{equation}
Let $\uvec_{1}\define \avec_{1}+\tilde{\Amat}_{1}\text{Vec}(\Gmat_{1})+\tilde{\Amat}_{2}\text{Vec}(\tilde{\Bmat}_{1})$ and $\uvec_{2} \define \avec_{1}+\tilde{\Amat}_{1}\text{Vec}(\Gmat_{2})+\tilde{\Amat}_{2}\text{Vec}(\tilde{\Bmat}_{2})$. Applying the triangle inequality, we obtain

\begin{equation}
    \begin{aligned}
        &\frac{1}{2}\|\alpha\uvec_{1} + (1-\alpha)\uvec_{2}\|_{2}^{2}\leq\frac{1}{2}(\alpha\|\uvec_{1}\|_{2}+(1-\alpha)\|\uvec_{2}\|_{2}|)^{2}.
    \end{aligned}
\end{equation}

It remains to prove that

\begin{equation}
    \begin{aligned}
        &\Bigg(\alpha\|\uvec_{1}\|_{2}+(1-\alpha)\|\uvec_{2}\|_{2}\Bigg)^{2}\leq \alpha\|\uvec_{1}\|_{2}^{2} + (1-\alpha)\|\uvec_{2}\|_{2}^{2}.
    \end{aligned}
\end{equation}
By expanding and simplifying the expression, we obtain

\begin{equation}
    \begin{aligned}
    \label{imp1}
    &\overbrace{(\alpha^{2}-\alpha)}^{\leq 0}\overbrace{\bigg(\|\uvec_{1}\|_{2}-\|\uvec_{2}\||_{2}\bigg)^{2}}^{\geq 0}\leq 0,
    \end{aligned}
\end{equation}
a condition that is always satisfied given $\alpha \in [0,1]$, since $\alpha^{2}-\alpha\leq 0$ and  $(\|\uvec_{1}\|_{2}-\|\uvec_{2}\||_{2})^{2}\geq 0$. 
\textbf{Q.E.D.}

\section{Derivation of the Hessian Matrix for Convexity Analysis}
\label{Hessian}
First, we note that according to \eqref{general_quadtratic_model} in the paper, the objective function satisfies
\begin{equation}
    \begin{aligned}  
   & \psi(\Xmat;\Gmat,\tilde{\Bmat})\\&\quad   = \frac{1}{2}\sum_{n=0}^{N-1}\|\avec_{1}+(\avec_{2}^{T}\otimes \Amat_{1})\text{Vec}(\Gmat)  +(\avec_{3}^{T}\otimes \Amat_{2})\text{Vec}(\tilde{\Bmat})\|_{2}^{2}
    \end{aligned}
\end{equation}
and
\begin{equation}
    \begin{aligned}
        &d\psi= \text{Trace}\Bigg(\Re\Bigg\{\sum_{n=0}^{N-1}\bigg(\avec_{1} + \tilde{\Amat}_{1}\text{Vec}(\Gmat)+\tilde{\Amat}_{2}\text{Vec}(\tilde{\Bmat})\bigg)^{H}\\&\quad \times\tilde{\Amat}_{1}\Bigg\}d\text{Vec}(\Gmat)\Bigg) \\&\quad + \text{Trace}\Bigg(\Re\Bigg\{\sum_{n=0}^{N-1}\bigg(\avec_{1} + \tilde{\Amat}_{1}\text{Vec}(\Gmat)+\tilde{\Amat}_{2}\text{Vec}(\tilde{\Bmat})\bigg)^{H}\\&\quad \times\tilde{\Amat}_{2}\Bigg\}d\text{Vec}(\tilde{\Bmat})\Bigg), 
    \end{aligned}
\end{equation}
where $\tilde{\Amat}_{1}\define \avec_{2}^{T}\otimes \Amat_{1}$ and $\tilde{\Amat}_{2} \define \avec_{3}^{T}\otimes \Amat_{2}$.

The first order derivatives of $\psi(\Xmat;\Gmat,\tilde{\Bmat})$ w.r.t. $\text{Vec}(\Gmat)$ and $\text{Vec}(\tilde{\Bmat})$ (denominator layout):
\begin{equation}
    \begin{aligned}
        &\nabla_{\text{Vec}(\Gmat)}\psi(\Xmat;\Gmat,\tilde{\Bmat})\define\frac{\partial\psi}{\partial \text{Vec}(\Gmat)}\\&\quad  = \sum_{n=0}^{N-1}\Re\Bigg\{\tilde{\Amat}_{1}^{H}\bigg(\avec_{1} + \tilde{\Amat}_{1}\text{Vec}(\Gmat)+\tilde{\Amat}_{2}\text{Vec}(\tilde{\Bmat})\bigg)\Bigg\},
    \end{aligned}
\end{equation}
\begin{equation}
    \begin{aligned}
        &\nabla_{\text{Vec}(\tilde{\Bmat})}\psi(\Xmat;\Gmat,\tilde{\Bmat})\define\frac{\partial\psi}{\partial \text{Vec}(\tilde{\Bmat})} \\&\quad = \sum_{n=0}^{N-1}\Re\Bigg\{\tilde{\Amat}_{2}^{H}\bigg(\avec_{1} + \tilde{\Amat}_{1}\text{Vec}(\Gmat)+\tilde{\Amat}_{2}\text{Vec}(\tilde{\Bmat})\bigg)\Bigg\}.
    \end{aligned}
\end{equation}

Now we calculate the second derivatives of $\psi(\Xmat;\Gmat,\tilde{\Bmat})$, as follows
\begin{equation}
    \begin{aligned}
        &\nabla_{\text{Vec}(\Gmat)}\bigg(\nabla_{\text{Vec}(\Gmat)}^{T}\psi(\Xmat;\Gmat,\tilde{\Bmat})\bigg) \\&\quad =\sum_{n=0}^{N-1} \Re\bigg\{\tilde{\Amat}_{1}^{T}\tilde{\Amat}_{1}^{*}\bigg\}\\&\quad =\sum_{n=0}^{N-1}\Re\bigg\{\bigg(\avec_{2}\avec_{2}^{H}\otimes \Amat_{1}^{T}\Amat_{1}^{*}\bigg)\bigg\},
    \end{aligned}
\end{equation}

\begin{equation}
    \begin{aligned}
        &\nabla_{\text{Vec}(\Gmat)}\bigg(\nabla_{\text{Vec}(\tilde{\Bmat})}^{T}\psi(\Xmat;\Gmat,\tilde{\Bmat})\bigg) \\&\quad = \sum_{n=0}^{N-1}\Re\bigg\{\tilde{\Amat}_{1}^{T}\tilde{\Amat}_{2}^{*}\bigg\}\\&\quad = \sum_{n=0}^{N-1}\Re\bigg\{\bigg(\avec_{2}\avec_{3}^{H}\otimes \Amat_{1}^{T}\Amat_{2}^{*}\bigg)\bigg\},
    \end{aligned}
\end{equation}

\begin{equation}
    \begin{aligned}
        &\nabla_{\text{Vec}(\tilde{\Bmat})}\bigg(\nabla_{\text{Vec}(\Gmat)}^{T}\psi(\Xmat;\Gmat,\tilde{\Bmat})\bigg)\\&\quad  = \sum_{n=0}^{N-1}\Re\bigg\{\tilde{\Amat}_{2}^{T}\tilde{\Amat}_{1}^{*}\bigg\} \\&\quad = \sum_{n=0}^{N-1}\Bigg(\Re\bigg\{\bigg(\avec_{2}\avec_{3}^{H}\otimes \Amat_{1}^{T}\Amat_{2}^{*}\bigg)\bigg\}\Bigg)^{T},
    \end{aligned}
\end{equation}

\begin{equation}
    \begin{aligned}
        &\nabla_{\text{Vec}(\tilde{\Bmat})}\bigg(\nabla_{\text{Vec}(\tilde{\Bmat})}^{T}\psi(\Xmat;\Gmat,\tilde{\Bmat})\bigg) \\&\quad = \sum_{n=0}^{N-1}\Re\bigg\{\tilde{\Amat}_{2}^{T}\tilde{\Amat}_{2}^{*}\bigg\} \\&\quad = \sum_{n=0}^{N-1}\Re\bigg\{\bigg(\avec_{3}\avec_{3}^{H}\otimes \Amat_{2}^{T}\Amat_{2}^{*}\bigg)\bigg\}.
    \end{aligned}
\end{equation}
Thus, we can conclude that the Hessian matrix is given by
\begin{equation}
    \begin{aligned}
        &\Hmat = \sum_{n=0}^{N-1} \\&\quad \begin{bmatrix}
            \Re\bigg\{(\avec_{2}\avec_{2}^{H}\otimes \Amat_{1}^{T}\Amat_{1}^{*})\bigg\} & \Re\bigg\{(\avec_{2}\avec_{3}^{H}\otimes \Amat_{1}^{T}\Amat_{2}^{*})\bigg\} \\ \Bigg(\Re\bigg\{(\avec_{2}\avec_{3}^{H}\otimes \Amat_{1}^{T}\Amat_{2}^{*})\bigg\}\Bigg)^{T}& \Re\bigg\{(\avec_{3}\avec_{3}^{H}\otimes \Amat_{2}^{T}\Amat_{2}^{*})\bigg\}
        \end{bmatrix},
    \end{aligned}
\end{equation}
which is a $2M^{2}\times 2M^{2}$ matrix.

\section{Proof of Positive Definiteness}
We will first show that the matrix $\Emat$ is positive semi-definite, which means it does not have any negative eigenvalues. After that, we explain why the matrix $\Emat$ is actually positive definite, meaning all its eigenvalues are positive. This two-step approach helps us confirm the positive nature of $\Emat$.
\subsubsection{\textbf{Proof of Positive Semi-Definiteness}}
Consider the matrix $\Emat$ defined as:
\begin{equation}
\begin{aligned}
&\Emat \define (\onevec_{M}\onevec_{M}^{T}\otimes \Imat_{M})  \\&\quad+ 2\Bigg(\Imat_{M^{2}} - \sum_{i=1}^{M}\sum_{j=1}^{M}\bigg((\evec_{i}\evec_{j}^{T})\otimes(\evec_{j}\evec_{i}^{T})\bigg)\Bigg) \\&\quad + \Lmat_{\ell}\Lmat_{\ell}^{T}.
\end{aligned}
\end{equation}
We aim to prove that $\Emat$ is a positive semi-definite matrix. The proof is structured as follows:

\paragraph{First Component Analysis}
For any vector $\alphavec \in \mathbb{R}^{M^{2}}$, we consider:
\begin{equation}
\begin{aligned}
\alphavec^{T}(\onevec_{M}\onevec_{M}^{T}\otimes \Imat_{M})\alphavec \geq 0.
\end{aligned}
\end{equation}
Utilizing the identity $(\Amat\otimes \Bmat)(\Cmat\otimes \Dmat) = (\Amat\Cmat)\otimes (\Bmat\Dmat)$ and the transpose property $(\Amat\otimes \Bmat)^{T} = \Amat^{T}\otimes \Bmat^{T}$, we derive:
\begin{equation}
\begin{aligned}
\alphavec^{T}(\onevec_{M}\otimes \Imat_{M})\overbrace{(\onevec_{M}\otimes \Imat_{M})^{T}\alphavec}^{\define \betavec} = \betavec^{T}\betavec = \|\betavec\|_{2}^{2} \geq 0.
\end{aligned}
\end{equation}

\paragraph{Analysis of the Identity and Commutation Matrix Difference}
We focus on the $\Imat_{M^{2}}$ and the commutation matrix difference. To prove this component is PSD:
\begin{equation}
\begin{aligned}
\Imat_{M^{2}} - \overbrace{\sum_{i=1}^{M}\sum_{j=1}^{M}\bigg((\evec_{i}\evec_{j}^{T})\otimes(\evec_{j}\evec_{i}^{T})\bigg)}^{\Kmat^{(M,M)}} \succeq \Zeromat_{M^{2}}.
\end{aligned}
\end{equation}
Given the property of the commutation matrix:
\begin{equation}
\begin{aligned}
\Kmat^{(M,M)}\Kmat^{(M,M),T} = \Kmat^{(M,M),T}\Kmat^{(M,M)} = \Imat_{M^{2}},
\end{aligned}
\end{equation}
it follows that all eigenvalues of $\Kmat^{(M,M)}\Kmat^{(M,M),T}$ are 1. Thus, the eigenvalues of $\Kmat^{(M,M)}$ span $\{-1, 1\}^{M^{2}}$, leading to eigenvalues of $\Imat_{M^{2}} - \Kmat^{(M,M)}$ being in $\{0, 2\}^{M^{2}}$. Consequently, the matrix is PSD.

\paragraph{Analysis of the $\Lmat_{\ell}$ Matrix}
The matrix $\Lmat_{\ell}$ is particularly structured to arrange elements strictly below the main diagonal of any matrix $\Amat \in \mathbb{R}^{M\times M}$ into $\text{Vec}(\Amat)$. Following a similar approach:
\begin{equation}
\begin{aligned}
\alphavec^{T}\Lmat_{\ell}\overbrace{\Lmat_{\ell}^{T}\alphavec}^{\define \betavec} = \betavec^{T}\betavec = \|\betavec\|_{2}^{2} \geq 0.
\end{aligned}
\end{equation}
Since the sum of PSD matrices remains PSD, $\Emat$ is conclusively positive semi-definite.\\
\textbf{Q.E.D.}
\subsubsection{\textbf{Positive Definiteness}}
Consider again the matrix $\Emat$ defined as: 
\begin{equation}
\begin{aligned}
&\Emat \define (\onevec_{M}\onevec_{M}^{T}\otimes \Imat_{M}) \\&\quad  + 2\left(\Imat_{M^{2}} - \sum_{i=1}^{M}\sum_{j=1}^{M}\left((\evec_{i}\evec_{j}^{T})\otimes(\evec_{j}\evec_{i}^{T})\right)\right) + \Lmat_{\ell}\Lmat_{\ell}^{T}.
\end{aligned}
\end{equation}
There exists $\forall M>0$ a positive real number $0<c<2$ such that:
\begin{equation}
    \begin{aligned}
        &c\cdot\Imat_{M^{2}}  + (\onevec_{M}\onevec_{M}^{T}\otimes \Imat_{M}) -2\Kmat^{(M,M)}+\Lmat_{\ell}\Lmat_{\ell}^{T}\succeq \Zeromat_{M^{2}}.
    \end{aligned}
\end{equation}
The matrix $(2-c)\Imat_{M^{2}}\succ\Zeromat_{M^{2}}$. The sum of a positive semi-definite matrix and a positive definite matrix results in a positive definite matrix. Consequently, the matrix $\Emat\succ \Zeromat_{M^{2}}$.


\end{document}